\RequirePackage{fix-cm} 
\RequirePackage[reqno]{amsmath}
\RequirePackage[l2tabu, orthodox]{nag}
\documentclass[a4paper, twoside, dvips, reqno, 12pt]{amsart}
\usepackage{etex}
\usepackage{fixltx2e}   

\usepackage[T1]{fontenc}
\usepackage[latin1]{inputenc}

\usepackage{esint}
\usepackage{amsbsy}
\usepackage{dsfont}
\usepackage{empheq}
\usepackage{xspace}
\usepackage{amsgen}
\usepackage{amsthm}
\usepackage{manfnt}
\usepackage{pifont}
\usepackage{amssymb}
\usepackage{upgreek}
\usepackage{amsfonts}
\usepackage{extarrows}
\usepackage{mathtools}
\usepackage{textgreek}
\usepackage[nice]{nicefrac}

\usepackage{mathabx}
\usepackage{relsize}
\usepackage{textcomp}
\usepackage[mathcal]{euscript}

\usepackage{rsfso}
\usepackage{mathrsfs}
\DeclareMathAlphabet{\mathscrbf}{OMS}{mdugm}{b}{n}

\DeclareFontFamily{U}{dutchcal}{\skewchar\font=45 }
\DeclareFontShape{U}{dutchcal}{m}{n}{<-> s*[1.0] dutchcal-r}{}
\DeclareFontShape{U}{dutchcal}{b}{n}{<-> s*[1.0] dutchcal-b}{}
\DeclareMathAlphabet{\mathlcal}{U}{dutchcal}{m}{n}
\SetMathAlphabet{\mathlcal}{bold}{U}{dutchcal}{b}{n}

\usepackage{a4wide}

\usepackage[dvipsnames, table]{xcolor}
\usepackage[most]{tcolorbox}

\definecolor{bckg}{RGB}{20.8, 20.8, 20.8}
\definecolor{oneblue}{rgb}{0.0, 0.0, 0.85}
\definecolor{Lightblue}{RGB}{214, 214, 214}
\definecolor{bluepigment}{rgb}{0.2, 0.2, 0.6}
\definecolor{charcoal}{rgb}{0.21, 0.27, 0.31}
\definecolor{denimblue}{rgb}{0.08, 0.38, 0.74}
\definecolor{Lightgray}{rgb}{0.89, 0.89, 0.89}
\definecolor{darkgrey}{rgb}{0.273, 0.281, 0.30}
\definecolor{darkelectricblue}{rgb}{0.33, 0.41, 0.47}

\usepackage[sort&compress, comma, square, numbers]{natbib}

\usepackage{tikz}
\usepackage{psfrag}
\usepackage{graphicx}
\usepackage{subfigure}
\usepackage{morefloats}
\usepackage{indentfirst}

\usepackage{acronym}
\usepackage{microtype}
\usepackage[labelsep=period,%
            labelfont={bf,sf,color=bluepigment},%
            justification=raggedright]{caption}

\usepackage[perpage, symbol]{footmisc}

\usepackage[usenames, dvipsnames, pdf]{pstricks}
\usepackage{pst-grad} 
\usepackage{pst-plot} 

\usepackage[colorlinks,
           urlcolor=oneblue,
           linkcolor=denimblue,
           citecolor=NavyBlue,
           bookmarksopen=false,
           pdfpagemode=UseNone,
           plainpages=false,
           pagebackref]{hyperref}

\usepackage[explicit]{titlesec}

\titleformat{\section}
  {\color{NavyBlue}\Large\sffamily\bfseries}
  {}
  {0em}
  {\colorbox{bckg!5}{\parbox{\dimexpr\linewidth-2\fboxsep\relax}{\centering\thesection. #1}}}
  [\vspace*{0.33em}]

\titleformat{name=\section, numberless}
  {\color{NavyBlue}\Large\sffamily\bfseries}
  {}
  {0.0em}
  {\colorbox{bckg!5}{\parbox{\dimexpr\linewidth-2\fboxsep\relax}{\centering#1}}}
  [\vspace*{0.33em}]

\titleformat{\subsection}
  {\color{NavyBlue}\large\sffamily\bfseries}
  {}
  {0.0em}
  {\colorbox{bckg!5}{\parbox{\dimexpr\linewidth-5\fboxsep\relax}{\centering\thesubsection. #1}}}
  [\vspace*{0.33em}]

\titleformat{name=\subsection, numberless}
  {\color{NavyBlue}\Large\sffamily\bfseries}
  {}
  {0em}
  {\colorbox{bckg!5}{\parbox{\dimexpr\linewidth-5\fboxsep\relax}{\centering#1}}}
  [\vspace*{0.33em}]

\titleformat{\subsubsection}
  {\color{bluepigment}\sffamily\normalsize\bfseries}
  {}
  {0.0em}
  {\colorbox{bckg!5}{\parbox{\dimexpr\linewidth-5\fboxsep\relax}{\centering\thesubsubsection. #1}}}
  [\vspace*{0.33em}]

\titleformat{name=\subsubsection, numberless}
  {\color{bluepigment}\sffamily\normalsize\bfseries}
  {}
  {0.0em}
  {\colorbox{bckg!5}{\parbox{\dimexpr\linewidth-5\fboxsep\relax}{\centering#1}}}
  [\vspace*{0.33em}]

\titleformat{\paragraph}[runin]
  {\color{bluepigment}\sffamily\small\bfseries}
  {}
  {0em}
  {#1}

\titlespacing{\section}{1.0em}{1.5em plus 2pt minus 2pt}%
{1.0em plus 2pt minus 2pt}[0em]
\titlespacing{\subsection}{1.0em}{1.5em plus 2pt minus 2pt}%
{1.0em}[0em]
\titlespacing{\subsubsection}{1.0em}{1.5em plus 2pt minus 2pt}%
{1.0em plus 2pt minus 2pt}[0em]

\usepackage{titletoc}

\contentsmargin{0.5em}
\newlength{\tocsep} 
\titlecontents{section}[\tocsep]
  {\addvspace{10pt}\bfseries\sffamily}
  {\contentslabel[\thecontentslabel]{\tocsep}}
  {}
  {\ \titlerule*[0.75pc]{.}\ \thecontentspage}
  []

\titlecontents{subsection}[\tocsep]
  {\addvspace{8pt}\sffamily}
  {\contentslabel[\thecontentslabel]{\tocsep}}
  {}
  {\ \titlerule*[0.5pc]{.}\ \thecontentspage}
  []

\titlecontents{subsubsection}[\tocsep]
  {\addvspace{4pt}\footnotesize\sffamily}
  {\contentslabel[\thecontentslabel]{\tocsep}\hspace*{0.5em}}
  {}
  {\ \titlerule*[0.35pc]{.}\ \thecontentspage}
  []

\makeatletter
\def\@setauthors{%
  \begingroup
  \def\thanks{\protect\thanks@warning}%
  \trivlist
  \centering\footnotesize \@topsep30\p@\relax
  \advance\@topsep by -\baselineskip
  \item\relax
  \author@andify\authors
  \def\\{\protect\linebreak}%
  \textsc{\normalsize\textcolor{darkelectricblue}{\authors}}%
  \ifx\@empty\contribs
  \else
    ,\penalty-3 \space \@setcontribs
    \@closetoccontribs
  \fi
  \endtrivlist
  \endgroup
}
\def\@settitle{\begin{center}%
  \baselineskip14\p@\relax
    \bfseries
    \textsc{\Large\textcolor{charcoal}{\@title}}
  \end{center}%
}
\makeatother

\usepackage{enumitem}
\setlist[description]{%
  topsep = 30pt,               
  itemsep = 8pt,               
  labelsep = 10pt,
  font={\bfseries\color{NavyBlue}}, 
}

\usepackage{fancyhdr}
\usepackage{lastpage}

\usepackage{fancyhdr}
\usepackage{lastpage}

\newcommand*\Title{\textcolor{bluepigment}{On the multi-symplectic structure of Boussinesq-type equations I}}
\newcommand*\Authors{\textcolor{bluepigment}{A.~Dur\'an, D.~Dutykh \& D.~Mitsotakis}}
\newcommand*{\plogo}{\textcolor{gray}{{\texttt{arXiv.org} / \textsc{hal}}}} 

\numberwithin{equation}{section}

\newtheorem{remark}{Remark}

\newcommand{\up}[1]{${}^{\,\mathrm{\textsf{#1}}}$} 

\newcommand{\R}{\mathds{R}}

\newcommand{\ui}{\mathrm{i}}

\newcommand{\eps}{\varepsilon}
\renewcommand{\O}{\mathcal{O}}
\renewcommand{\nu}{\text{\textnu}}

\renewcommand{\eta}{\text{\texteta}}
\renewcommand{\beta}{\text{\textbeta}}
\renewcommand{\mu}{\text{\textmugreek}}
\renewcommand{\alpha}{\text{\textalpha}}
\renewcommand{\kappa}{\text{\textkappa}}
\renewcommand{\omega}{\text{\textomega}}
\renewcommand{\theta}{\text{\texttheta}}
\newcommand{\ud}{\mathrm{d}\hspace{0.08em}}

\newcommand{\J}{\mathds{J}}
\newcommand{\K}{\mathds{K}}
\newcommand{\M}{\mathds{M}}
\newcommand{\DD}{\mathrm{D}}
\newcommand{\T}{\mathcal{T}}
\newcommand{\Pp}{\mathcal{P}}
\newcommand{\Ss}{\mathcal{S}}
\newcommand{\A}{\mathfrak{A}}
\newcommand{\B}{\mathfrak{B}}
\newcommand{\E}{\mathfrak{E}}
\newcommand{\F}{\mathfrak{F}}
\newcommand{\I}{\mathfrak{I}}
\newcommand{\Ec}{\mathcal{E}}
\newcommand{\Ic}{\mathcal{I}}
\newcommand{\Gg}{\mathfrak{G}}
\newcommand{\Gm}{\mathfrak{M}}
\newcommand{\Rr}{\mathfrak{R}}
\renewcommand{\H}{\mathscr{H}}
\renewcommand{\geq}{\geqslant}
\renewcommand{\leq}{\leqslant}
\newcommand{\z}{\boldsymbol{z}}
\renewcommand{\S}{\mathfrak{S}}

\newcommand{\Rn}{\mathscr{R}}
\newcommand{\D}{\mathfrak{D}}
\newcommand{\Ll}{\mathscr{L}}
\renewcommand{\L}{\mathcal{L}}
\newcommand{\Lin}{\mathfrak{L}}
\renewcommand{\mapsto}{\longmapsto}

\DeclareMathOperator{\rank}{rank}

\newcommand{\cf}{\emph{cf.}\xspace}
\newcommand{\ie}{\emph{i.e.}\xspace}
\newcommand{\eg}{\emph{e.g.}\xspace}

\newcommand{\etal}{\emph{et al.}\xspace}

\newcommand{\sech}{\mathrm{sech}}

\newcommand{\scal}{\boldsymbol{\cdot}}
\newcommand{\grad}{\boldsymbol{\nabla}}

\newcommand{\abs}[1]{\lvert\, #1\, \rvert}
\newcommand{\sign}{\mathop{\mathrm{sign}}}

\newcommand{\pd}[2]{\frac{\partial #1}{\partial\/ #2}}

\newcommand{\Prod}[2]{\left\langle\, #1,\,#2\,\right\rangle}

\newcommand{\eqdef}{\mathop{\stackrel{\,\mathrm{def}}{:=}\,}}

\newcommand{\vdd}[2]{\dfrac{\delta #1}{\delta\hspace{0.0556em} #2}}

\newcommand{\half}{{\textstyle{1\over2}}}
\newcommand{\sixth}{{\textstyle{1\over6}}}
\renewcommand{\third}{{\textstyle{1\over3}}}
\renewcommand{\fourth}{{\textstyle{1\over4}}}
\newcommand{\threefourth}{{\textstyle{3\over4}}}

\makeatletter
\renewcommand*\env@matrix[1][\arraystretch]{%
  \edef\arraystretch{#1}%
  \hskip -\arraycolsep
  \let\@ifnextchar\new@ifnextchar
  \array{*\c@MaxMatrixCols c}}
\makeatother

\usepackage{acronym}
\acrodef{bvp}[BVP]{Boundary Value Problem}
\acrodef{NSWE}{Nonlinear Shallow Water Equations}

\begin{document}

\title[\Title]{On the multi-symplectic structure of Boussinesq-type systems. I: Derivation and mathematical properties}

\author[A.~Dur\'an]{Angel Dur\'an}
\address{\textbf{A.~Dur\'an:} Departamento de Matem\'atica Aplicada, E.T.S.I. Telecomunicaci\'on, Campus Miguel Delibes, Universidad de Valladolid, Paseo de Belen 15, 47011 Valladolid, Spain}
\email{angel@mac.uva.es\vspace*{0.5em}}

\author[D.~Dutykh]{Denys Dutykh$^*$}
\address{\textbf{D.~Dutykh:} Univ. Grenoble Alpes, Univ. Savoie Mont Blanc, CNRS, LAMA, 73000 Chamb\'ery, France and LAMA, UMR 5127 CNRS, Universit\'e Savoie Mont Blanc, Campus Scientifique, F-73376 Le Bourget-du-Lac Cedex, France}
\email{Denys.Dutykh@univ-savoie.fr}
\urladdr{http://www.denys-dutykh.com/\vspace*{0.5em}}
\thanks{$^*$ Corresponding author}

\author[D.~Mitsotakis]{Dimitrios Mitsotakis}
\address{\textbf{D.~Mitsotakis:} Victoria University of Wellington, School of Mathematics and Statistics, PO Box 600, Wellington 6140, New Zealand}
\email{dmitsot@gmail.com}
\urladdr{http://dmitsot.googlepages.com/}



\begin{titlepage}
\clearpage
\pagenumbering{arabic}
\thispagestyle{empty} 
\noindent
{\Large Angel \textsc{Dur\'an}}\\
{\it\textcolor{gray}{Universidad de Valladolid, Spain}}
\\[0.02\textheight]
{\Large Denys \textsc{Dutykh}}\\
{\it\textcolor{gray}{CNRS, Universit\'e Savoie Mont Blanc, France}}
\\[0.02\textheight]
{\Large Dimitrios \textsc{Mitsotakis}}\\
{\it\textcolor{gray}{Victoria University of Wellington, New Zealand}}
\\[0.16\textheight]

\vspace*{0.99cm}

\colorbox{Lightblue}{
  \parbox[t]{1.0\textwidth}{
    \centering\huge\sc
    \vspace*{0.75cm}
    
    \textcolor{bluepigment}{On the multi-symplectic structure of Boussinesq-type systems. I: Derivation and mathematical properties}
    
    \vspace*{0.75cm}
  }
}

\vfill 

\raggedleft     
{\large \plogo} 
\end{titlepage}


\thispagestyle{empty} 
\par\vspace*{\fill}   
\begin{flushright} 
{\textcolor{denimblue}{\textsc{Last modified:}} \today}
\end{flushright}


\begin{abstract}

The \textsc{Boussinesq} equations are known since the end of the XIX\up{st} century. However, the proliferation of various \textsc{Boussinesq}-type systems started only in the second half of the XX\up{st} century. Today they come under various flavours depending on the goals of the modeller. At the beginning of the XXI\up{st} century an effort to classify such systems, at least for even bottoms, was undertaken and developed according to both different physical regimes and mathematical properties, with special emphasis, in this last sense, on the existence of symmetry groups and their connection to conserved quantities. Of particular interest are those systems admitting a symplectic structure, with the subsequent preservation of the total energy represented by the \textsc{Hamiltonian}. In the present paper a family of \textsc{Boussinesq}-type systems with multi-symplectic structure is introduced. Some properties of the new systems are analyzed: their relation with already known \textsc{Boussinesq} models, the identification of those systems with additional \textsc{Hamiltonian} structure as well as other mathematical features like well-posedness and existence of different types of solitary-wave solutions. The consistency of multi-symplectic systems with the full \textsc{Euler} equations is also discussed.


\bigskip\bigskip
\noindent \textbf{\keywordsname:} multi-symplectic structure; long dispersive wave; Boussinesq equations; surface waves \\

\smallskip \\
\noindent \textbf{PACS [2010]:} 47.35.Bb (primary), 47.35.Fg (secondary)

\end{abstract}


\newpage
\maketitle
\thispagestyle{empty}


\clearpage
\thispagestyle{empty}
\tableofcontents


\clearpage
\vspace*{0.25em}
\section{Introduction}

The first \textsc{Boussinesq}-type equation was proposed by Joseph \textsc{Boussinesq} in 1877 \cite{Boussinesq1877}, who gave the name to different families of nonlinear wave equations proposed since then. This topic remained a \emph{sleeping beauty}, \cite{Ke2015}, during almost one hundred years. The modern era was opened by Howell \textsc{Peregrine} in his theoretical and numerical investigations of long wave transformations on uniform slopes \cite{Peregrine1967}. A proliferation of various systems started, serving various purposes of the near-shore hydrodynamics. Some reviews of this topic can be found in \cite{Madsen1999, DMII, Brocchini2013}.

\begin{figure}
  \centering
  \includegraphics[width=0.79\textwidth]{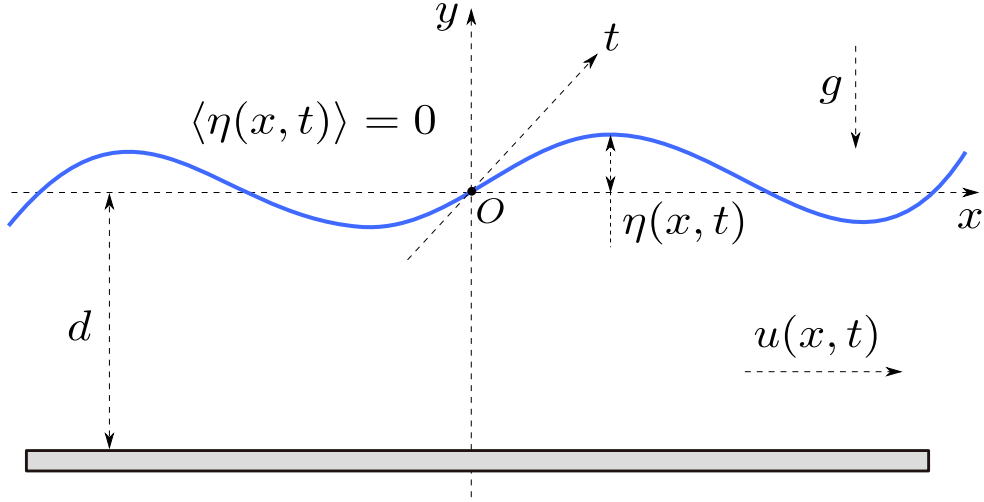}
  \caption{\small\em Sketch of the fluid domain. We note that it is always possible to choose dimensionless variables such that $g\ =\ 1$ and $d\ =\ 1\,$. That is why these constants are absent in governing Equations \eqref{eq:1}, \eqref{eq:2} (such a change of dependent and independent variables is given in \cite[Equation~(3.13)]{DMII}). The brackets $\langle\scal\rangle$ denote the spatial averaging operator, which fix the still water level in our problem.}
  \label{fig:sketch}
\end{figure}

At the beginning of the XXI\up{st} century an effort to classify such systems, at least for even bottoms, was undertaken. In this way, the following four-parameter family 
\begin{align}\label{eq:0}
  \eta_{\,t}\ +\ [\,u\ +\ \eta\,u\,]_{\,x}\ +\ a\,u_{\,x\,x\,x}\ -\ b\,\eta_{\,x\,x\,t}\ &=\ 0\,, \\
  u_{\,t}\ +\ \bigl[\,\eta\ +\ \half\;u^{\,2}\,\bigr]_{\,x}\ +\ c\,\eta_{\,x\,x\,x}\ -\ d\,u_{\,x\,x\,t}\ &=\ 0\,,\label{eq:00}
\end{align}
denoted here as the  $(a\,,\,b\,,\,c\,,\,d)$ family, was formulated and analyzed by \textsc{Bona} \etal, \cite{BCS, Bona2004}. In \eqref{eq:0}, \eqref{eq:00}, $\eta\ =\ \eta\,(x,\,t)$ and $u\ =\ u\,(x,\,t)$ are real-valued functions defined for $x\ \in\ \R$ and $t\ \geq\ 0\,$, while the coefficients $a,\,b,\,c,\,d$ are defined as
\begin{align}\label{eq:0b}
  a\ &\eqdef\ \half\;\bigl(\theta^{\,2}\ -\ \third\bigr)\,\nu\,, \qquad
  & b\ \eqdef\ \half\;\bigl(\theta^{\,2}\ -\ \third\bigr)\cdot(1\ -\ \nu)\,, \\
  c\ &\eqdef\ \half\;\bigl(1\ -\ \theta^{\,2}\bigr)\,\mu\,, \qquad
  & d\ \eqdef\ \half\;\bigl(1\ -\ \theta^{\,2}\bigr)\cdot(1\ -\ \mu)\,,\label{eq:00b}
\end{align}
where $\nu\,$, $\mu\ \in\ \R$ and $0\ \leq\ \theta\ \leq\ 1\,$. The Systems \eqref{eq:0}, \eqref{eq:00} are proposed for modelling the two-way propagation of one-dimensional, small amplitude, irrotational long surface waves in a channel of constant depth, see Figure~\ref{fig:sketch}. The variables $x$ and $t$ represent, respectively, the position along the channel and time, while  $\eta\,(x,\,t)$ and $u\,(x,\,t)$ are proportional to the free surface excursion and to the horizontal velocity of the fluid at $(x,\,t)$ at a non-dimensional height $y\ =\ -1\ +\ \theta\,(1\ +\ \eta\,(x,\,t))\,$, respectively. A review of the existing mathematical theory for the Systems \eqref{eq:0}, \eqref{eq:00} which includes a list of systems of the family of especial interest, well-posedness results of the corresponding Initial -- Value Problem (IVP) and periodic IVP as well as existence and stability results of solitary-wave solutions, can be seen in \cite{DMII}. A symmetric\footnote{Here, the symmetry is understood in the sense of \textsc{Friedrichs}--\textsc{Lax} \cite{Friedrichs1971}.} variant of the $(a\,,\,b\,,\,c\,,\,d)$ system of the general form
\begin{align}\label{eq:1}
  \eta_{\,t}\ +\ [\,u\ +\ \half\;\eta\,u\,]_{\,x}\ +\ a\,u_{\,x\,x\,x}\ -\ b\,\eta_{\,x\,x\,t}\ &=\ 0\,, \\
  u_{\,t}\ +\ \bigl[\,\eta\ +\ \fourth\;\eta^{\,2}\ +\ \threefourth\;u^{\,2}\,\bigr]_{\,x}\ +\ c\,\eta_{\,x\,x\,x}\ -\ d\,u_{\,x\,x\,t}\ &=\ 0\,,\label{eq:2}
\end{align}
with $a,\,b,\,c,\,d$ as in \eqref{eq:0b}, \eqref{eq:00b}, was proposed by \textsc{Bona} \etal in \cite{BCL}. These systems retain the same order of approximation to the \textsc{Euler} equations as \eqref{eq:0}, \eqref{eq:00}, reduce to a symmetric hyperbolic system when the dispersive third-order derivative terms are omitted, preserve the $L^{\,2}$ norm
\begin{align}\label{eq:l2norm}
  \int_{\,-\,\infty}^{\,+\,\infty}(\,\eta^{\,2}\ +\ u^{\,2}\,)\;\ud x\,,
\end{align}
and, concerning existence and uniqueness of solutions of the corresponding IVP, are locally well-posed. Higher-order \textsc{Boussinesq}-type systems are proposed in \eg \cite{BCS} and also by \textsc{Daripa} in \cite{Daripa2006}. Although we focus here on surface wave theory, it may be worth mentioning the derivation of \textsc{Boussinesq} systems for internal waves in, \eg \cite{BLS2008}.

It is well-known that only a sub-family of $(a\,,\,b\,,\,c\,,\,d)$ systems \eqref{eq:0}, \eqref{eq:00} with $b\ =\ d$ possesses a co-symplectic \textsc{Hamiltonian} structure
\begin{equation}\label{eq:ham}
  \begin{pmatrix}
    \eta_{\,t} \\
    u_{\,t}
  \end{pmatrix}\ =\ 
  \J\scal\begin{pmatrix}[2]
    \vdd{\H}{\eta} \\
    \vdd{\H}{u}
  \end{pmatrix}\,,
\end{equation}
on a suitable functional space for $(\eta\,,\, u)\,$, where the co-symplectic structure is given by the nonlocal matrix operator
\begin{equation}\label{eq:sympl}
  \J\ \eqdef\ \begin{pmatrix}
  0 & -\,\bigl(1\ -\ b\,\partial_{\,x\,x}^{\,2}\bigr)^{\,-1}\circ\partial_{\,x} \\
  -\,\bigl(1\ -\ b\,\partial_{\,x\,x}^{\,2}\bigr)^{\,-1}\circ\partial_{\,x} & 0
  \end{pmatrix}\,,
\end{equation}
(the symbol $\circ$ denotes composition) and $\H$ is the \textsc{Hamiltonian} functional given by
\begin{equation*}
  \H\ \eqdef\ \frac{1}{2}\;\int_{\,\R}\bigl\{\,\eta^{\,2}\ +\ (1\ +\ \eta\,)\,u^{\,2}\ -\ c\,\eta_{\,x}^{\,2}\ -\ a\,u_{\,x}^{\,2}\,\bigr\}\;\ud x\,.
\end{equation*}
with $\vdd{\H}{\eta}\,$, $\vdd{\H}{u}$ standing for the variational (\ie \textsc{Fr\'echet}) derivatives of $\H$ with respect to $\eta$ and $u\,$, respectively. (This \textsc{Hamiltonian} structure is lost in the symmetric version \eqref{eq:1}, \eqref{eq:2}.) Note that the co-symplectic operator $\J$ does not depend explicitly neither on time $t\,$, nor on space $x\,$, nor on the solution $(\eta\,,\,u)\,$. In this sense the symplectic structure is quite rigid for Partial Differential Equations (PDEs). For each IVP of \eqref{eq:ham}, the \textsc{Hamiltonian} functional $\H\,(t)$ is conserved in time and its value, determined by the corresponding initial conditions, can be considered as a generalized `energy' of the state of the system represented by the solution, \cite{Christov2001}. In addition to $\H\,$, the impulse functional
\begin{align*}\label{eq:h1norm}
  \int_{\,-\,\infty}^{\,+\,\infty}(\,\eta \, u\ +b\ u_{x}\, \eta_{x}\,)\;\ud x\,,
\end{align*}
is preserved by smooth, decaying enough at infinity solutions.

The present paper delves into the properties of \textsc{Boussinesq}-type systems focusing on the multi-symplectic (MS) structure. It is the first part of a study devoted by the authors to this question and deals with the derivation of a MS family of  equations of \textsc{Boussinesq} type, the comparison with other existing Boussinesq systems and the description of some mathematical properties, with special emphasis on the existence of traveling-wave solutions. The second part of the study, devoted to the construction of multi-symplectic schemes for the approximation to the systems, will be the subject of a forthcoming paper.

The multi-symplectic theory generalizes the classical \textsc{Hamiltonian} formulations, \cite{Basdevant2007}, to the case of PDEs such that the space and time variables are treated on the equal footing \cite{Bridges1997} (see also \cite[Chapter~12]{Leimkuhler2004}). Multi-symplectic formulations are also gaining popularity for both mathematical investigations, \cite{Marsden1998}, along with numerical \emph{structure preserving} modeling, \cite{Bridges2001, Dutykh2013a}.

The history of multi-symplectic formulations can be traced back to V.~\textsc{Volterra} (1890) who generalized \textsc{Hamiltonian} equations for variational problems involving several variables, \cite{Volterra1890, Volterra1890a}. Later these ideas were developed in 1930's, \cite{DeDonder1930, Weyl1935, Lepage1936}. Finally, in 1970's this theory was geometrized by several mathematical physicists, \cite{Goldschmidt1973, Kijowki1974, Krupka1975, Krupka1975a} similarly to the evolution of symplectic geometry from the ideas of J.-L.~\textsc{Lagrange}, \cite{Lagrange1853, Souriau1997}. In our study we will be inspired by modern works on multi-symplectic PDEs, \cite{Bridges1997, Marsden1998}. Recently this theory has found many applications to the development of structure-preserving integrators with different strategies, \cite{Bridges2001, Moore2003a, Chen2011, Dutykh2013a}.

Let us also briefly review the main known equations arising in the modeling of long waves with MS formulation. The KdV equation is among the multi-symplectic veterans, \cite{Zhao2000, Bridges2001}, along with the NLS equation, \cite{Chen2002}. Among one-way propagation models, the multi-symplectic structure of the \textsc{Benjamin--Bona--Mahony} (BBM) and generalized\footnote{The generalization consisted in taking higher order nonlinearities, \ie $u\,u_{\,x}\ \leadsto\ u^{\,q}\,u_{\,x}\,$, $q\ \geq\ 1\,$.} BBM equations was highlighted in \cite{Sun2004}. Some interesting numerical results for the BBM equation were presented in \cite{Li2013}. The `good' scalar \textsc{Boussinesq} equation was studied in \cite{Huang2003}. The multi-symplectic structure of the celebrated \textsc{Serre}--\textsc{Green}--\textsc{Naghdi} (SGN) equations was unveiled in \cite{Chhay2016}. Finally, the multi-symplectic structure of two-layer SGN system modeling the propagation of long fully nonlinear internal waves was highlighted recently in \cite{Clamond2016a}.  We also remind that the full water wave problem is multi-symplectic, \cite{Bridges1997}.

The main highlights of the present paper revolve around the derivation of a family of \textsc{Boussinesq} systems with multi-symplectic structure. The procedure for the derivation starts from a system of $(a\,,\,b\,,\,c\,,\,d)$ form with a general homogeneous quadratic polynomial as nonlinear term and determines a combination of parameters allowing the multi-symplectic formulation. Additional advantages of the technique implemented are the identification of the MS structure in \textsc{Boussinesq} systems of the family \eqref{eq:1}, \eqref{eq:2} and the way how to modify the Equations~\eqref{eq:0}, \eqref{eq:00} to admit the MS property.

After the derivation, it is expected to discuss some properties of the new systems. In this sense, the present paper is focused on the identification of those equations of the family with additional symplectic structure, the well-posedness of the corresponding IVP and the existence of different types of solitary-wave solutions. For the last two points, the form of the nonlinearity enables us to make use of the existing literature on them. The discussion of consistency of the \textsc{Euler} equations with the new system is here initiated from the MS structure of Equations~\eqref{eq:1}, \eqref{eq:2}. Considered as part of this discussion are also some comparisons of the solitary-wave speed-amplitude relations for the MS systems with systems~\eqref{eq:0}, \eqref{eq:00} and the \textsc{Euler} system. These comparisons suggest the existence of new MS systems, derived in this paper, with good performance in this sense. Nevertheless, a complete analysis of the question deserves a future research. The description of the models is completed with the construction and development of multi-symplectic schemes of approximation, being  the subject of a second, forthcoming part.

The paper is organized according to the following structure. After a brief reminder of MS theory for PDEs, Section~\ref{sec:2} is devoted to the procedure of derivation of the MS structure. The main steps are first described to obtain the MS formulations of the KdV--BBM equation and the symmetric System \eqref{eq:1}, \eqref{eq:2}. The technique is then extended to construct a family of MS \textsc{Boussinesq}-type systems and to explain how to modify \eqref{eq:0}, \eqref{eq:00} in such a way that the resulting systems are multi-symplectic. Some properties of the new equations concerning well-posedness and existence of solitary-wave solutions are studied in Section~\ref{sec:sec3}. The main conclusions and perspectives are outlined in Section~\ref{sec:sec4}.


\section{Multi-symplectic structure}
\label{sec:2}

We recall first some basic facts about multi-symplectic geometry and PDEs. A PDE\footnote{Note that there is no assumption that the IVP for this PDE is well-posed. To give an example, the multi-symplectic setting may include some elliptic PDEs in space-time which are ill-posed in the sense of \textsc{Hadamard}.} (or a system of PDEs) is said to be multi-symplectic in one space dimension $x\ \in\ \R$ if it can be written in the following canonical form:
\begin{equation}\label{eq:ms}
  \K\scal\z_{\,t}\ +\ \M\scal\z_{\,x}\ =\ \grad_{\,\z}\,\S\,(\z)\,, \qquad \z\ \in\ \R^{\,d},
\end{equation}
for some $d\ \geq\ 3\,$, where $\z\,(x,\,t)\,:\ \R\times\R^{\,+}\ \mapsto\ \R^{\,d}\,$, $\K\ \in\ \mathcal{M}_{\,d\,\times\,d}\,(\R)$ and $\M\ \in\ \mathcal{M}_{\,d\,\times\,d}\,(\R)$ are some real, {skew-symmetric} $d\times d$ matrices, the dot $\scal$ denotes the matrix-vector product in $\R^{\,d}$ and $\grad_{\,\z}$ is the classical gradient operator in $\R^{\,d}\,$. The given function $\S\,(\z)$ is assumed to be a smooth function of its $d$ variables $\z\ =\ \bigl(z_{\,1},\,z_{\,2},\,\ldots,z_{\,d}\bigr)\,$.

The formulation \eqref{eq:ms} represents a direct generalization of \textsc{Hamiltonian} PDEs where the space and time are treated on equal footing. On the other hand, the main drawback of formulation \eqref{eq:ms} is that it is not intrinsic, \ie not coordinate-free on the base manifold $(x,\,t)\ \in\ \R\times\R^{\,+}\,$. However, it is sufficient for our purposes, since in wave propagation we would like to keep the distinction between space and time coordinates.

All equations with the multi-symplectic form \eqref{eq:ms}, have also a certain number of relevant properties. One is the existence of conservation laws. Note first that Equations~\eqref{eq:ms} satisfy the \emph{multi-symplectic conservation law}:
\begin{equation}\label{eq:cons}
  \omega_{\,t}\ +\ \kappa_{\,x}\ =\ 0\,,
\end{equation}
with
\begin{equation}\label{eq:def}
  \omega\ \eqdef\ \frac{1}{2}\;\ud\z\,\wedge\,(\K\scal\ud\z)\,, \qquad
  \kappa\ \eqdef\ \frac{1}{2}\;\ud\z\,\wedge\,(\M\scal\ud\z)\,,
\end{equation}
with $\wedge$ being the standard exterior product of differential forms, \cite{Spivak1971}. The $2-$form $\omega$ defines a symplectic structure on $\R^{\,\rank{\K}}\,$ (where $\rank{\K}$ denotes the rank of the matrix $\K$), which is associated with the time direction and the $2-$form $\kappa$ defines a symplectic structure $\R^{\,\rank{\M}}\,$, which is associated with the space direction, \cite{Bridges2001}. From definitions of $2-$forms \eqref{eq:def}, we can see that the multi-symplectic approach relies on a local concept of symplecticity, since symplectic forms may vary with the solution in space and in time similarly to a \textsc{Lagrangian} tracer in a fluid flow. This observation explains why the class of multi-symplectic PDEs is more general. On the other hand, when designing a geometric integrator, the principal requirement is that the discretization conserves exactly the symplecticity \eqref{eq:cons}.

\begin{remark}
In classical field theories sometimes the following $3-$form is called a multi-symplectic form:
\begin{equation*}
 \boldsymbol{\Omega}\ \eqdef\ \omega\,\wedge\,\ud x\ -\ \kappa \,\wedge\,\ud t\ -\ \ud\S\,\wedge\,\ud x\,\wedge\,\ud t\ \in\ \bigwedge^{3}\,\bigl(\,\R^{\,d}\times\R^{\,2}\,\bigr)\,.
\end{equation*}
In \cite{Bridges2001} it was proposed to call $\boldsymbol{\Omega}$ a \emph{meta-symplectic form} and we follow this terminology. Throughout our manuscript the term \emph{multi-symplectic form} refers to the system of PDEs \eqref{eq:ms} or to the couple of differential forms $\omega,\ \kappa \ \in\ \bigwedge^{2}\,\bigl(\,\R^{\,d}\,\bigr)$ depending on the context.
\end{remark}

Additionally, other conservation laws can be derived when the function $\S\,(\z)$ does not depend explicitly on $x$ or $t\,$. Defining the generalized energy $\E$ and generalized momentum $\I$ densities as
\begin{equation*}
  \E\,(\z)\ \eqdef\ \S\,(\z)\ -\ \half\;\Prod{\z}{\M\scal\z_{\,x}}\,, \qquad
  \I\,(\z)\ \eqdef\ \half\;\Prod{\z}{\K\scal\z_{\,x}}\,,
\end{equation*}
and corresponding fluxes
\begin{equation*}
  \F\,(\z)\ \eqdef\ \half\;\Prod{\z}{\M\scal\z_{\,t}}\,, \qquad
  \Gm\,(\z)\ \eqdef\ \S\,(\z)\ -\ \half\;\Prod{\z}{\K\scal\z_{\,t}}\,,
\end{equation*}
(the symbol $\Prod{\cdot}{\cdot}$ denotes the standard scalar product in $\R^{\,d}\,$) then the generalized energy $\Ec\ \eqdef\ \int_{\,\R}\,\E\,(\z)\,\ud x$ and momentum $\Ic\ \eqdef\ \int_{\,\R}\,\I\,(\z)\,\ud x$ are conserved:
\begin{equation}\label{eq:cons2}
  \E_{\,t}\ +\ \F_{\,x}\ =\ 0\,, \qquad \I_{\,t}\ +\ \Gm_{\,x}\ =\ 0\,,
\end{equation}
Note that one of the advantages of the multi-symplectic structure is that a geometric interpretation of the conservation laws can be derived from the application of \textsc{Noether} theory. Note also that, in general, it is not possible to conserve energy and momentum \eqref{eq:cons2} with \emph{uniform discretizations} additionally to multi-symplecticity \eqref{eq:cons}.

Finally, multi-symplectic PDEs \eqref{eq:ms} automatically possess also the \textsc{Lagrangian} variational structure. Indeed, it is not difficult to check that Equation~\eqref{eq:ms} is the \textsc{Euler}--\textsc{Lagrange} equation to the following \textsc{Lagrangian} functional:
\begin{equation*}
  \L\ =\ \int_{\,t_{\,0}}^{\,t_{\,1}}\Ll\,\ud t\,, \qquad \L\ \rightharpoonup\ \mathrm{St}\,,
\end{equation*}
with the \textsc{Lagrangian} density defined as
\begin{equation*}
  \Ll\ \eqdef\ \int_{\,x_{\,0}}^{\,x_{\,1}}\Bigl\{\half\;\Prod{\K\scal\z_{\,t}}{\z}\ +\ \half\;\Prod{\M\scal\z_{\,x}}{\z}\ -\ \S\,(\z)\Bigr\}\,\ud x\,.
\end{equation*}
Here $t_{\,0}\,$, $t_{\,1}$ denote arbitrary instances of time $t_{\,1}\ >\ t_{\,0}\ \geq\ 0$ and, similarly, $x_{\,0}\,$, $x_{\,1}$ are some locations in space $x_{\,1}\ >\ x_{\,0}\,$. From the last definition, observe that the function $\S\,(\z)$ plays the r\^ole of the generalized potential energy. The \textsc{Hamilton} principle states that the physical trajectory corresponds to stationary values of the action functional, \ie $\L\ \rightharpoonup\ \mathrm{St}\,$.


\subsection{MS structure of the KdV--BBM equation}
\label{sec:kb}

As a necessary step towards understanding the way how the multi-symplectic structure of the $(a\,,\,b\,,\,c\,,\,d)-$type systems will be derived later, we first study the case of the KdV--BBM equation 
\begin{equation}\label{eq:kb}
  u_{\,t}\ +\ u\,u_{\,x}\ +\ \alpha\,u_{\,x\,x\,x}\ -\ \beta\,u_{\,x\,x\,t}\ =\ 0\,,
\end{equation}
where $\alpha\,$, $\beta\ \in\ \R$ are some real coefficients. The variable $u\,(x,\,t)$ can be related both to the velocity or to the free surface elevation. The KdV--BBM Equation~\eqref{eq:kb} arises in the modeling of unidirectional water wave propagation \cite{Dutykh2010e} and in the study of non-integrable solitonic gases \cite{Dutykh2014d}. It was also used to study the well-posedness of the KdV equation in \cite{Bona1975a}.) The multi-symplectic structure of \eqref{eq:kb} has never been reported before to the best of our knowledge. We provide below an example of such structure in $\R^{\,5}\,$.

At the first step, we rewrite Equation~\eqref{eq:kb} in a conservative form:
\begin{equation}\label{eq:ckb}
  u_{\,t}\ +\ \Bigl[\,\half\;u^{\,2}\ +\ \alpha\,u_{\,x\,x}\ -\ \beta\,u_{\,x\,t}\,\Bigr]_{\,x}\ =\ 0\,.
\end{equation}
In order to lower the order of the equation, the following variables are introduced:
\begin{equation*}
  \phi_{\,x}\ \eqdef\ u\,, \qquad
  v\ \eqdef\ u_{\,x}\,, \qquad
  w\ \eqdef\ u_{\,t}\,.
\end{equation*}
The new variable $\phi$ is a generalized potential for the field $u\,$. This potential appears, for example, in \textsc{Lagrangian} variational formulations of both KdV and BBM equations. As the final conceptual step, we rewrite Equation~\eqref{eq:ckb} in the following slightly unusual way:
\begin{equation*}
  \half\;u_{\,t}\ +\ \Bigl[\,\underbrace{\half\;u^{\,2}\ +\ \half\;\phi_{\,t}\ +\ \alpha\,v_{\,x}\ -\ \half\;\beta\,w_{\,x}\ -\ \half\;\beta\,v_{\,t}}_{\equiv\ p}\,\Bigr]_{\,x}\ =\ 0\,.
\end{equation*}
The last equation suggests the introduction of an extra variable, which has the meaning of a flux:
\begin{equation*}
  p\ \eqdef\ \half\;u^{\,2}\ +\ \half\;\phi_{\,t}\ +\ \alpha\,v_{\,x}\ -\ \half\;\beta\,w_{\,x}\ -\ \half\;\beta\,v_{\,t}\,.
\end{equation*}
Now, we have all elements to present the desired multi-symplectic structure. The vector $\z$ including the field $u$ along with `conjugate momenta' is defined as
\begin{equation*}
  \z\ \eqdef\ \bigl(\,u\,,\,\phi\,,\,v\,,\,w\,,\,p\,\bigr)\ \in\ \R^{\,5}\,.
\end{equation*}
Note then that \eqref{eq:kb} is equivalent to the system of equations:
\begin{align}\label{eq:ms1}
  \half\;\phi_{\,t}\ -\ \half\;\beta\,v_{\,t}\ +\ \alpha\,v_{\,x}\ -\ \beta\,w_{\,x}\ &=\ p\ -\ \half\;u^{\,2}\,, \\
  -\half\;u_{\,t}\ -\ p_{\,x}\ &=\ 0\,, \label{eq:ms2} \\
  \half\;\beta\,u_{\,t}\ -\ \alpha\,u_{\,x}\ &=\ -\alpha\,v\ +\ \half\;\beta\,w\,, \label{eq:ms3} \\
  \half\;\beta\,u_{\,x}\ &=\ \half\;\beta\,v\,, \label{eq:ms4} \\
  \phi_{\,x}\ &=\ u\,. \label{eq:ms5}
\end{align}
We now consider the vector field $\Rr\,:\ \R^{\,5}\ \mapsto\ \R^{\,5}\,$ given by the right hand side of \eqref{eq:ms1} -- \eqref{eq:ms5}:
\begin{equation*}
  \Rr\,(\z)\ \eqdef\ {}^{\top}\,\Bigl(\,p\ -\ \half\;u^{\,2}\,, 0\,,\ -\alpha\,v\ +\ \half\;\beta\,w\,, \half\;\beta\,v\,, u\,\Bigr)\,.
\end{equation*}
Since the \textsc{Jacobian} ${\Rr}^{\prime}\,(\z)$ of $\Rr$ is symmetric for all $\z\,$, then \textsc{Poincar\'e} lemma implies that $\Rr$ is conservative and thus $\Rr\ =\ \grad\,\S$ for some potential $\S\,$. Therefore, System \eqref{eq:ms1} -- \eqref{eq:ms5} can be written in the canonical matrix-vector form \eqref{eq:ms} with the following skew-symmetric matrices:
\begin{equation*}
  \K\ =\ \begin{pmatrix}[1.1]
    0 & \half & -\half\;\beta & 0 & 0 \\
    -\half & 0 & 0 & 0 & 0 \\
    \half\;\beta & 0 & 0 & 0 & 0 \\
    0 & 0 & 0 & 0 & 0 \\
    0 & 0 & 0 & 0 & 0
  \end{pmatrix}\,, \qquad
  \M\ =\ \begin{pmatrix}[1.1]
    0 & 0 & \alpha & -\half\;\beta & 0 \\
    0 & 0 & 0 & 0 & -1 \\
    -\alpha & 0 & 0 & 0 & 0 \\
    \half\;\beta & 0 & 0 & 0 & 0 \\
    0 & 1 & 0 & 0 & 0
  \end{pmatrix}\,.
\end{equation*}
A potential energy $\S\,(\z)$ for the KdV--BBM Equation~\eqref{eq:kb} is
\begin{equation*}
  \S\,(\z)\ =\ p\,u\ -\ \sixth\;u^{\,3}\ -\ \half\;\alpha\,v^{\,2}\ +\ \half\;\beta\,v\,w\,.
\end{equation*}
This completes the definition of the multi-symplectic structure for \eqref{eq:kb}. We conjecture that another multi-symplectic structure for this equation in $\R^{\,d}\,$, with $d\ <\ 5\,$ is not possible.


\subsection{MS structure of the symmetric $(a\,,\,b\,,\,c\,,\,d)$ Boussinesq system}
\label{sec:ms}

The previous steps can be adapted to obtain the MS structure of some systems of the family \eqref{eq:1}, \eqref{eq:2}. Similarly to Section~\ref{sec:kb}, these can first be written in the conservative form
\begin{align}\label{eq:3s}
  \eta_{\,t}\ +\ [\,u\ +\ \half\;\eta\,u\ +\ a\,u_{\,x\,x}\ -\ b\,\eta_{\,x\,t}\,]_{\,x}\ &=\ 0\,, \\
  u_{\,t}\ +\ \bigl[\,\eta\ +\ \fourth\;\eta^{\,2}\ +\ \threefourth\;u^{\,2}\ +\ c\,\eta_{\,x\,x}\ -\ d\,u_{\,x\,t}\,\bigr]_{\,x}\ &=\ 0\,,\label{eq:4s}
\end{align}
and we introduce the additional variables:
\begin{equation}\label{eq:3sb}
  \phi_{\,1\,x}\ \eqdef\ \eta\,, \qquad
  v_{\,1}\ \eqdef\ \eta_{\,x}\,, \qquad
  w_{\,1}\ \eqdef\ \eta_{\,t}\,,
\end{equation}
\begin{equation}\label{eq:4sb}
  \phi_{\,2\,x}\ \eqdef\ u\,, \qquad
  v_{\,2}\ \eqdef\ u_{\,x}\,, \qquad
  w_{\,2}\ \eqdef\ u_{\,t}\,,
\end{equation}
where $\phi_{\,1,\,2}$ are generalized potentials and $v_{\,1,\,2}\,$, $w_{\,1,\,2}$ are space and time gradients of dynamic variables $\eta$ and $u$ correspondingly. Using these variables, we can rewrite Equations~\eqref{eq:3s}, \eqref{eq:4s} as
\begin{align*}
  \half\;\eta_{\,t}\ +\ \Bigl[\,\underbrace{u\ +\ \half\;\eta\,u\ +\ \half\;\phi_{\,1\,t}\ +\ a\,v_{\,2\,x}\ -\ \half\;b\,v_{\,1\,t}\ -\ \half\;b\,w_{\,1\,x}}_{\equiv\ p_{\,1}}\,\Bigr]_{\,x}\ &=\ 0\,, \\
  \half\;u_{\,t}\ +\ \Bigl[\,\underbrace{\eta\ +\ \fourth\;\eta^{\,2}\ +\ \threefourth\;u^{\,2}\ +\ \half\;\phi_{\,2\,t}\ +\ c\,v_{\,1\,x}\ -\ \half\;d\,v_{\,2\,t}\ -\ \half\;d\,w_{\,2\,x}}_{\equiv\ p_{\,2}}\,\Bigr]_{\,x}\ &=\ 0\,.
\end{align*}
Thus, defining two additional fluxes:
\begin{align}\label{eq:3sbb}
  p_{\,1}\ &\eqdef\ u\ +\ \half\;\eta\,u\ +\ \half\;\phi_{\,1\,t}\ +\ a\,v_{\,2\,x}\ -\ \half\;b\,v_{\,1\,t}\ -\ \half\;b\,w_{\,1\,x}\,, \\
  p_{\,2}\ &\eqdef\ \eta\ +\ \fourth\;\eta^{\,2}\ +\ \threefourth\;u^{\,2}\ +\ \half\;\phi_{\,2\,t}\ +\ c\,v_{\,1\,x}\ -\ \half\;d\,v_{\,2\,t}\ -\ \half\;d\,w_{\,2\,x}\,\label{eq:4sbb},
\end{align}
then the vector $\z$ with auxiliary variables can now be defined:
\begin{equation}\label{eq:zdef}
  \z\ \eqdef\ \bigl(\,\eta\,,\,\phi_{\,1},\,v_{\,1},\,w_{\,1},\,p_{\,1},\,u,\,\phi_{\,2},\,v_{\,2},\,w_{\,2},\,p_{\,2}\,\bigr)\ \in\ \R^{\,10}\,.
\end{equation}
System~\eqref{eq:3s}, \eqref{eq:4s} is given below in the expanded form:
\begin{align*}
  \half\;\phi_{\,1\,t}\ -\ \half\;b\,v_{\,1\,t}\ +\ a\,v_{\,2\,x}\ -\ \half\;b\,w_{\,1\,x}\ &=\ p_{\,1}\ -\ u\ -\ \half\;\eta\,u\,, \\
  -\,\half\;\eta_{\,t}\ -\ p_{\,1\,x}\ &=\ 0\,, \\
  \half\;b\,\eta_{\,t}\ -\ a\,u_{\,x}\ &=\ \half\;b\,w_{\,1}\ -\ a\,v_{\,2}\,, \\
  \half\;b\,\eta_{\,x}\ &=\ \half\;b\,v_{\,1}\,, \\
  \phi_{\,1\,x}\ &=\ \eta\,, \\
  \half\;\phi_{\,2\,t}\ -\ \half\;d\,v_{\,2\,t}\ +\ c\,v_{\,1\,x}\ -\ \half\;d\,w_{\,2\,x}\ &=\ p_{\,2}\ -\ \eta\ -\ \fourth\;\eta^{\,2}\ -\ \threefourth\;u^{\,2}\,, \\
  -\,\half\;u_{\,t}\ -\ p_{\,2\,x}\ &=\ 0\,, \\
  \half\;d\,u_{\,t}\ -\ a\,\eta_{\,x}\ &=\ \half\;d\,w_{\,2}\ -\ c\,v_{\,1}\,, \\
  \half\;d\,u_{\,x}\ &=\ \half\;d\,v_{\,2}\,, \\
  \phi_{\,2\,x}\ &=\ u\,.
\end{align*}
Now, if $\Rr\,:\ \R^{\,10}\ \mapsto\ \R^{\,10}$ is the vector field whose components are given by the right hand side:
\begin{multline*}
  \Rr\,(\z)\ \eqdef\ {}^{\top}\,\Bigl(\,p_{\,1}\ -\ u\ -\ \half\;\eta\,u\,,\, 0\,,\, \half\;b\,w_{\,1}\ -\ a\,v_{\,2}\,,\, \half\;b\,v_{\,1},\,\eta\,, \\
  p_{\,2}\ -\ \eta\ -\ \fourth\;\eta^{\,2}\ -\ \threefourth\;u^{\,2}\,,\, 0\,,\, \half\;d\,w_{\,2}\ -\ c\,v_{\,1}\,,\, \half\;d\,v_{\,2}\,,\, u\,\Bigr)\,,
\end{multline*}
then we observe that the \textsc{Jacobian} ${\Rr}^{\,\,\prime}\,(\z)$ of $\Rr$ is symmetric for all $\z$ if and only if $a\ =\ c\,$. Under this condition, the last system of scalar equations can be recast into the canonical matrix-vector form \eqref{eq:ms}, if we introduce the following skew-symmetric matrices:
\begin{equation}\label{eq:K}
  \K\ =\ \begin{pmatrix}[1.1]
    0 & \half & -\half\;b & 0 & 0 & 0 & 0 & 0 & 0 & 0 \\
    -\half & 0 & 0 & 0 & 0 & 0 & 0 & 0 & 0 & 0 \\
    \half\;b & 0 & 0 & 0 & 0 & 0 & 0 & 0 & 0 & 0 \\
    0 & 0 & 0 & 0 & 0 & 0 & 0 & 0 & 0 & 0 \\
    0 & 0 & 0 & 0 & 0 & 0 & 0 & 0 & 0 & 0 \\
    0 & 0 & 0 & 0 & 0 & 0 & \half & -\half\;d & 0 & 0 \\
    0 & 0 & 0 & 0 & 0 & -\half & 0 & 0 & 0 & 0 \\
    0 & 0 & 0 & 0 & 0 & \half\;d & 0 & 0 & 0 & 0 \\
    0 & 0 & 0 & 0 & 0 & 0 & 0 & 0 & 0 & 0 \\
    0 & 0 & 0 & 0 & 0 & 0 & 0 & 0 & 0 & 0
  \end{pmatrix}\,,
\end{equation}
\begin{equation}\label{eq:M}
  \M\ =\ \begin{pmatrix}[1.1]
    0 & 0 & 0 & -\half\;b & 0 & 0 & 0 & a & 0 & 0 \\
    0 & 0 & 0 & 0 & -1 & 0 & 0 & 0 & 0 & 0 \\
    0 & 0 & 0 & 0 & 0 & -c & 0 & 0 & 0 & 0 \\
    \half\;b & 0 & 0 & 0 & 0 & 0 & 0 & 0 & 0 & 0 \\
    0 & 1 & 0 & 0 & 0 & 0 & 0 & 0 & 0 & 0 \\
    0 & 0 & c & 0 & 0 & 0 & 0 & 0 & -\half\;d & 0 \\
    0 & 0 & 0 & 0 & 0 & 0 & 0 & 0 & 0 & -1 \\
    -a & 0 & 0 & 0 & 0 & 0 & 0 & 0 & 0 & 0 \\
    0 & 0 & 0 & 0 & 0 & \half\;d & 0 & 0 & 0 & 0 \\
    0 & 0 & 0 & 0 & 0 & 0 & 1 & 0 & 0 & 0
  \end{pmatrix}\,,
\end{equation}
and where a potential energy for the symmetric $(a\,,\,b\,,\,a\,,\,d)$ sub-family reads:
\begin{equation*}
  \S\,(\z)\ \eqdef\ p_{\,1}\,\eta\ -\ \eta\,u\ -\ \fourth\;\eta^{\,2}\,u\ +\ \half\;b\,v_{\,1}\,w_{\,1}\ -\ \fourth\;u^{\,3}\ +\ \half\;d\,v_{\,2}\,w_{\,2}\ -\ a\,v_{\,1}\,v_{\,2}\ +\ p_{\,2}\,u\,.
\end{equation*}
This completes the presentation of the multi-symplectic structure of Equations~\eqref{eq:3s}, \eqref{eq:4s}. It is noted that only a sub-class of symmetric Systems \eqref{eq:1}, \eqref{eq:2} (with $a\ =\ c$) possesses a multi-symplectic structure.


\subsection{A MS family of Boussinesq-type systems}
\label{sec:ms2}

The construction presented in the previous sections can be conformed to derive a new family of \textsc{Boussinesq}-type systems with MS structure. We start from a system of equations of the general form
\begin{align}\label{eq:5s}
  \eta_{\,t}\ +\ [\,u\ +\ \A\,(\eta\,,\,u)\ +\ a\,u_{\,x\,x}\ -\ b\,\eta_{\,x\,t}\,]_{\,x}\ &=\ 0\,, \\
  u_{\,t}\ +\ \bigl[\,\eta\ +\ \B\,(\eta\,,\,u)\ +\ c\,\eta_{\,x\,x}\ -\ d\,u_{\,x\,t}\,\bigr]_{\,x}\ &=\ 0\,, \label{eq:6s}
\end{align}
of \textsc{Boussinesq} type. The parameters $(a,\, b,\, c,\, d)$ are as in \eqref{eq:0b}, \eqref{eq:00b} and the nonlinearities $\A$ and $\B$ are homogeneous, quadratic polynomials
\begin{align*}
  \A\,(\eta\,,\,u)\ &\eqdef\ \alpha_{\,1\,1}\,\eta^{\,2}\ +\ \alpha_{\,1\,2}\,\eta\,u\ +\ \alpha_{\,2\,2}\,u^{\,2}\,, \\
  \B\,(\eta\,,\,u)\ &\eqdef\ \beta_{\,1\,1}\,\eta^{\,2}\ +\ \beta_{\,1\,2}\,\eta\,u\ +\ \beta_{\,2\,2}\,u^{\,2}\,,
\end{align*}
with real coefficients $\alpha_{\,\imath\,\jmath}\,$, $\beta_{\,\imath\,\jmath}$ which can be chosen on modeling or geometric structure bases. System of the form \eqref{eq:5s}, \eqref{eq:6s} have been used for modelling nonlinear waves in different situations. Several examples are given below:
\begin{enumerate}
  \item The $(a\,,\,b\,,\,c\,,\,d)$ System \eqref{eq:0}, \eqref{eq:00} trivially corresponds to the choice
  \begin{align*}
  \alpha_{\,1\,1}\ &=\ 0\,, \quad \alpha_{\,1\,2}\ =\ 1\,, \quad \alpha_{\,2\,2}\ =\ 0\,, \\
  \beta_{\,1\,1}\ &=\ 0\,, \quad \beta_{\,1\,2}\ =\ 0\,, \quad \beta_{\,2\,2}\ =\ \frac{1}{2}\,.
  \end{align*}
  \item The symmetric version \eqref{eq:1}, \eqref{eq:2} is obtained from \eqref{eq:5s}, \eqref{eq:6s} with 
  \begin{align*}
  \alpha_{\,1\,1}\ &=\ 0\,, \quad \alpha_{\,1\,2}\ =\ \frac{1}{2}\,, \quad \alpha_{\,2\,2}\ =\ 0\,, \\
  \beta_{\,1\,1}\ &=\ \frac{1}{4}\,, \quad \beta_{\,1\,2}\ =\ 0\,, \quad \beta_{\,2\,2}\ =\ \frac{3}{4}\,.
  \end{align*}
  \item Some examples of \eqref{eq:5s}, \eqref{eq:6s} also appear by choosing specific values of $(a\,,\,b\,,\,c\,,\,d)$ instead of the parameters of the nonlinearities. This is the case of the system of KdV type considered by \textsc{Bona} \etal in \cite{BonaChK}:
  \begin{align}\label{eq:5sb}
    \eta_{\,t}\ +\ [\ \A\,(\eta\,,\,u)\ +\ u_{\,x\,x}\,]_{\,x}\ &=\ 0\,, \\
    u_{\,t}\ +\ \bigl[\ \B\,(\eta\,,\,u)\ +\ \eta_{\,x\,x}\,\bigr]_{\,x}\ &=\ 0\,, \label{eq:6sb}
  \end{align}
  with $\A$ and $\B$ as above, which corresponds to taking $a\ =\ c\ =\ 1\,$, $b\ =\ d\ =\ 0$ in \eqref{eq:5s}, \eqref{eq:6s} and where the terms $u_{\,x}$ and $\eta_{\,x}$ can be omitted by a suitable change of variables. 
  \item Other examples can be found in the \textsc{Gear}--\textsc{Grimshaw} system, \cite{GearG}, for internal wave propagation or in the coupled systems of BBM-type considered in \cite{Hakkaev}.
\end{enumerate}

In order to obtain a MS structure in the Equations~\eqref{eq:5s}, \eqref{eq:6s}, we consider \eqref{eq:3sb}, \eqref{eq:4sb} and generalize \eqref{eq:3sbb}, \eqref{eq:4sbb} by introducing  the auxiliary variables 
\begin{align}\label{eq:5sc}
  p_{\,1}\ &\eqdef\ u\ +\ \A\,(\eta\,,\,u)\ +\ \half\;\phi_{\,1\,t}\ +\ a\,v_{\,2\,x}\ -\ \half\;b\,v_{\,1\,t}\ -\ \half\;b\,w_{\,1\,x}\,, \\
  p_{\,2}\ &\eqdef\ \eta\ +\ \B\,(\eta\,,\,u)\ +\ \half\;\phi_{\,2\,t}\ +\ c\,v_{\,1\,x}\ -\ \half\;d\,v_{\,2\,t}\ -\ \half\;d\,w_{\,2\,x}\,.\label{eq:6sc}
\end{align}
Then, Equations~\eqref{eq:5s}, \eqref{eq:6s} can be written in an equivalent form as the following system of first-order differential relations:
\begin{align*}
  \half\;\phi_{\,1\,t}\ -\ \half\;b\,v_{\,1\,t}\ +\ a\,v_{\,2\,x}\ -\ \half\;b\,w_{\,1\,x}\ &=\ p_{\,1}\ -\ u\ -\ \A\,(\eta\,,\,u)\,, \\
  -\,\half\;\eta_{\,t}\ -\ p_{\,1\,x}\ &=\ 0\,, \\
  \half\;b\,\eta_{\,t}\ -\ a\,u_{\,x}\ &=\ \half\;b\,w_{\,1}\ -\ a\,v_{\,2}\,, \\
  \half\;b\,\eta_{\,x}\ &=\ \half\;b\,v_{\,1}\,, \\
  \phi_{\,1\,x}\ &=\ \eta\,, \\
  \half\;\phi_{\,2\,t}\ -\ \half\;d\,v_{\,2\,t}\ +\ c\,v_{\,1\,x}\ -\ \half\;d\,w_{\,2\,x}\ &=\ p_{\,2}\ -\ \eta\ -\ \B\,(\eta\,,\,u)\,, \\
  -\,\half\;u_{\,t}\ -\ p_{\,2\,x}\ &=\ 0\,, \\
  \half\;d\,u_{\,t}\ -\ a\,\eta_{\,x}\ &=\ \half\;d\,w_{\,2}\ -\ c\,v_{\,1}\,, \\
  \half\;d\,u_{\,x}\ &=\ \half\;d\,v_{\,2}\,, \\
  \phi_{\,2\,x}\ &=\ u\,.
\end{align*}
Now the vector field $\Rr\,(\z)$ (with vector $\z$ defined as in \eqref{eq:zdef}) on the right hand side is of the form
\begin{multline*}
  \Rr\,(\z)\ \eqdef\ {}^{\top}\,\Bigl(\,p_{\,1}\ -\ u\ -\ \A\,(\eta\,,\,u),\, 0,\,\half\;b\,w_{\,1}\ -\ a\,v_{\,2},\,\half\;b\,v_{\,1},\,\eta,\\
  p_{\,2}\ -\ \eta\ -\ \B\,(\eta\,,\,u),\,0,\,\half\;d\,w_{\,2}\ -\ a\,v_{\,1},\,\half\;d\,v_{\,2},\,u\,\Bigr)\,.
\end{multline*}
Imposing the symmetry of the \textsc{Jacobian} leads to
\begin{equation*}
  a\ =\ c\,, \qquad \pd{\,\A}{u}\ =\ \pd{\,\B}{\eta}\,.
\end{equation*}
The last condition holds for any values of $\eta$ and $u$ if and only if
\begin{equation}\label{eq:const}
  \alpha_{\,1\,2}\ =\ 2\,\beta_{\,1\,1}\,, \qquad
  \beta_{\,1\,2}\ =\ 2\,\alpha_{\,2\,2}\,.
\end{equation}
Henceforth, by \textsc{Poincar\'e} lemma, $\Rr\,(\z)$ is conservative, \ie $\Rr\,(\z)\ =\ \grad_{\,\z}\,\S\,(\z)\,$, when $a\ =\ c$ and \eqref{eq:const} hold. All generalized Systems~\eqref{eq:5s}, \eqref{eq:6s} with such coefficients are multi-symplectic with a potential functional given \eg by
\begin{multline*}
  \S\,(\z)\ \eqdef\ p_{\,1}\,\eta\ -\ \eta\,u\ -\ \third\;\alpha_{\,1\,1}\,\eta^{\,3}\ -\ \beta_{\,1\,1}\,\eta^{\,2}\,u\ -\ \half\;\beta_{\,1\,2}\,\eta\,u^{\,2}\ +\ \half\;b\,v_{\,1}\,w_{\,1} \\ 
  -\ \third\;\beta_{\,2\,2}\,u^{\,3}\ +\ \half\;d\,v_{\,2}\,w_{\,2}\ -\ a\,v_{\,1}\,v_{\,2}\ +\ p_{\,2}\,u\,.
\end{multline*}
The skew-symmetric matrices $\K$ and $\M$ are defined as in \eqref{eq:K}, \eqref{eq:M}.

\begin{remark}
It may be worth describing the application of this result to the above mentioned particular cases:
\begin{enumerate}
  \item From conditions~\eqref{eq:const} it can be readily seen that the asymptotically consistent $(a\,,\,b\,,\,a\,,\,d)$ family \eqref{eq:0}, \eqref{eq:00} is not multi-symplectic. Indeed,
  \begin{align*}
    1\ &=\ \alpha_{\,1\,2}\ \neq\ 2\,\beta_{\,1\,1}\ =\ 0\,, \\
    0\ &=\ \beta_{\,1\,2}\ =\ 2\,\alpha_{\,2\,2}\ =\ 0\,.
  \end{align*}
  From these observations it follows that the $(a\,,\,b\,,\,a\,,\,d)$ family \eqref{eq:0}, \eqref{eq:00} can be made naturally multi-symplectic with \emph{minimal} modifications if we take $\beta_{\,1\,1}\ =\ \frac{1}{2}\,$. All other coefficients are kept unchanged. The resulting multi-symplectic \textsc{Boussinesq}-type system reads:
  \begin{align}
    \eta_{\,t}\ +\ [\,u\ +\ \eta\,u\,]_{\,x}\ +\ a\,u_{\,x\,x\,x}\ -\ b\,\eta_{\,x\,x\,t}\ &=\ 0\,,\label{eq:a} \\
    u_{\,t}\ +\ \bigl[\,\eta\ +\ \underbrace{\half\;\eta^{\,2}}_{\displaystyle{(*)}}\ +\ \half\;u^{\,2}\,\bigr]_{\,x}\ +\ c\,\eta_{\,x\,x\,x}\ -\ d\,u_{\,x\,x\,t}\ &=\ 0\,.\label{eq:b}
  \end{align}
  The new term $(*)$ is asymptotically small, since its magnitude, in non-dimensional, scaled variables, is $\O\,(\eps^{\,2})\,$, where $\eps$ is the nonlinearity order parameter. 

  \item It is clear that the MS structure of the symmetric family of \eqref{eq:1}, \eqref{eq:2}, shown in Section~\ref{sec:ms}, can be alternatively obtained from the verification of the conditions derived here.

  \item In the Systems~\eqref{eq:5sb}, \eqref{eq:6sb}, since $a\ =\ c\ =\ 1\,$, the MS structure holds for those equations for which the coupled nonlinear terms satisfy \eqref{eq:const}. Furthermore, it can be seen that the conditions \eqref{eq:const} are related to the reduces system in \cite{BonaChK} to simplify the study of existence and stability of the solitary wave solutions.
\end{enumerate}
\end{remark}

\begin{remark}
Conditions $a\ =\ c\ \,$, and \eqref{eq:const} for a multi-symplectic structure of Systems~\eqref{eq:5s}, \eqref{eq:6s} were obtained by using the set of variables \eqref{eq:3sb}, \eqref{eq:4sb} and \eqref{eq:5sc}, \eqref{eq:6sc}. The derivation of other conditions with different variables or with extensions of the definition in \eqref{eq:ms} which allows the matrices $\K$ and $\M$ be dependent of $\z$, cf. \cite{Bridges1997}, is not discarded. Thus, the \textsc{Boussinesq} System~\eqref{eq:0b}, \eqref{eq:00b} with $a\ =\ 0\,$, $b\ =\ d\ =\ 1/3$ and $c\ =\ -1/3$ is multi-symplectified in \cite{Bridges1999}.
\end{remark}


\subsubsection{Systems with symplectic and multi-symplectic structure}

We note that if $b\ =\ d\,$, the Equations~\eqref{eq:5s}, \eqref{eq:6s} have a  \textsc{Hamiltonian} structure \eqref{eq:ham}, with the structure operator $\J$ as in \eqref{eq:sympl}, when the coefficients of the nonlinear terms $\A\,(\eta\,,\,u)$ and $\B\,(\eta\,,\,u)\,$ satisfy
\begin{equation}\label{eq:const2}
  \beta_{\,1\,2}\ =\ 2\,\alpha_{\,1\,1}\,, \qquad
  \alpha_{\,1\,2}\ =\ 2\,\beta_{\,2\,2}\,.
\end{equation}
The \textsc{Hamiltonian} is now given by
\begin{equation*}
  \H\ \eqdef\ \frac{1}{2}\;\int_{\,\R}\bigl\{\,\eta^{\,2}\ +\ u^{\,2}\ -\ c\,\eta_{\,x}^{\,2}\ -\ a\,u_{\,x}^{\,2}\ +\ 2\,\Gg\,(\eta\,,\,u)\,\bigr\}\;\ud x\,,
\end{equation*}
where
\begin{equation*}
  \Gg\,(\eta\,,\,u)\ \eqdef\ \frac{\beta_{\,1\,1}}{3}\;\eta^{\,3}\ +\ \frac{\beta_{\,1\,2}}{2}\;\eta^{\,2}\,u\ +\ \beta_{\,2\,2}\,\eta\,u^{\,2}\ +\ \frac{\alpha_{\,2\,2}}{3}\;u^{\,3}\,.
\end{equation*}
This allows for studying whether the intersection of symplectic and multi-symplectic \textsc{Boussinesq}-type Systems in \eqref{eq:5s}, \eqref{eq:6s} is non-empty. It is not hard to see that the compatibility of the conditions \eqref{eq:const} and \eqref{eq:const2} holds when
\begin{equation}\label{eq:const3}
  \beta_{\,1\,2}\ =\ 2\,\alpha_{\,1\,1}\ =\ 2\,\alpha_{\,2\,2}\,, \qquad
  \alpha_{\,1\,2}\ =\ 2\,\beta_{\,1\,1}\ =\ 2\,\beta_{\,2\,2}\,,
\end{equation}
and then the corresponding $(a\,,\,b\,,\,a\,,\,b)$ System~\eqref{eq:5s}, \eqref{eq:6s} is multi-symplectic and symplectic. The family is determined by two dispersive parameters (say $(a\,$, $b)$) and two nonlinear parameters (say $(\,\beta_{\,1\,1}\,$, $\beta_{\,1\,2}\,)$). The corresponding \textsc{Hamiltonian} is given by:
\begin{equation*}
  \H\ \eqdef\ \frac{1}{2}\;\int_{\,\R}\bigl\{\,\eta^{\,2}\ +\ u^{\,2}\ -\ a\,(\eta_{\,x}^{\,2}\ +\ u_{\,x}^{\,2})\ +\ 2\,\Gg\,(\eta\,,\,u)\,\bigr\}\;\ud x\,,
\end{equation*}
where
\begin{equation*}
  \Gg\,(\eta\,,\,u)\ \eqdef\ \frac{\beta_{\,1\,1}}{3}\;\eta^{\,3}\ +\ \frac{\beta_{\,1\,2}}{2}\;\eta^{\,2}\,u\ +\ \beta_{\,1\,1}\,\eta\,u^{\,2}\ +\ \frac{\beta_{\,1\,2}}{6}\;u^{\,3}\,.
\end{equation*}
Note that the family~\eqref{eq:1}, \eqref{eq:2} does not contain any system with both structures. (Actually, as mentioned before, any of them does not have a \textsc{Hamiltonian} formulation.) On the other hand, systems of KdV type  \eqref{eq:5sb}, \eqref{eq:6sb} (or of BBM type, \cite{Hakkaev}) with homogeneous quadratic nonlinearities, are symplectic and multi-symplectic when \eqref{eq:const3} holds. Finally, it can be seen that System~\eqref{eq:a}, \eqref{eq:b} with $b\ =\ d$ and $a\ =\ c$ is symplectic and multi-symplectic (for $\,\beta_{\,1\,1}\ =\ \frac{1}{2}\,$, $\beta_{\,1\,2}\ =\ 0\,$).


\section{Some properties of the MS Boussinesq-type systems}
\label{sec:sec3}

In this Section some additional properties of the MS \textsc{Boussinesq} Systems~\eqref{eq:5s}, \eqref{eq:6s} satisfying $a\ =\ c$ and \eqref{eq:const} will be discussed. They will be focused on the well-posedness of the corresponding IVP and the existence of solitary wave solutions. The results are mainly based on the literature about these questions for $(a\,,\,b\,,\,c\,,\,d)$ systems.

\subsection{Linear well-posedness}

Linear well-posedness can be studied using the arguments considered in \cite{BCS} for the $(a\,,\,b\,,\,c\,,\,d)$ System~\eqref{eq:0}, \eqref{eq:00}, since the linear part coincides with that of Equations~\eqref{eq:5s}, \eqref{eq:6s}. According to this, and if 
\begin{equation*}
  \omega_{\,1}\,(k)\ \eqdef\ \frac{1\ -\ a\,k^{\,2}}{1\ +\ b\,k^{\,2}}\,, \quad 
  \omega_{\,2}\,(k)\ \eqdef\ \frac{1 - c\,k^{\,2}}{1\ +\ d\,k^{\,2}}\,, \quad 
  k\ \in\ \R\,,
\end{equation*}
then the linearized problem is well-posed when the rational function $\omega_{\,1}\,(k)/\omega_{\,2}\,(k)$ has neither zeros or poles on the real axis. This limits the range of the parameters $(a\,,\,b\,,\,c\,,\,d)$ to the three `admissible' cases derived in \cite[Proposition~3.1]{BCS}. The MS structure, which requires the condition $a\ =\ c\,$, reduces them to
\begin{description}
  \item[$(L_{\,1})\ $]\qquad $b\ \geq\ 0\,$, \qquad $d\ \geq\ 0\,$, \qquad $a\ =\ c\ ,$
  \item[$(L_{\,2})\ $]\qquad $b\ =\ d\ <\ 0\,$, \qquad $a\ =\ c\ >\ 0$.
\end{description}
Furthermore, for $(a\,,\,b\,,\,c\,,\,d)$ satisfying one of the conditions $(L_{\,1})\,$, $(L_{\,2})\,$, if $\ell$ denotes the order of the symbol
\begin{equation*}
  g\,(k)\ \eqdef\ \sqrt{\frac{\omega_{\,1}\,(k)}{\omega_{\,2}\,(k)}}\,, \qquad k\ \in\ \R\,,
\end{equation*}
and $m_{\,1}\ \eqdef\ \max\,\{\,0,\,-\,\ell\,\}\,$, $m_{\,2}\ \eqdef\ \max\,\{\,0,\,\ell\,\}\,$, then \cite[Theorem~3.2]{BCS} applies to have linear well-posedness of the IVP in the $L^{\,2}$ based \textsc{Sobolev} spaces $H^{\,s\,+\,m_{\,1}}\,\times\,H^{\,s\,+\,m_{\,2}}$ for any $s\ \geq\ 0\,$.


\subsection{Nonlinear well-posedness}

The study of local well-posedness of the full nonlinear MS System~\eqref{eq:5s}, \eqref{eq:6s} may follow the lines established in \cite{Bona2004} for the IVP of the $(a\,,\,b\,,\,c\,,\,d)$ System~\eqref{eq:0}, \eqref{eq:00}. This is due to the fact that, although the nonlinear terms may be different, they are always homogeneous quadratic polynomials and, therefore, arguments and estimates used in \cite{Bona2004} can be taken in the new systems. This leads to the following cases of local well-posedness in suitable \textsc{Sobolev} spaces:
\begin{description}
  \item[$(N_{\,1})\ $]\qquad $a\ =\ c\,$, \qquad $b\ >\ 0\,$, \qquad $d\ >\ 0$.
  \item[$(N_{\,2})\ $]\qquad $a\ =\ c\ >\ 0\,$, \qquad $b\ =\ d\ =\ 0$ \quad (\cf \cite{BonaChK}),
  \item[$(N_{\,3})\ $]\qquad $a\ = c\,$, \qquad $b\ =\ 0\,$, \qquad $d\ >\ 0$.
  \item[$(N_{\,4})\ $]\qquad $a\ =\ c\ \geq\ 0\,$, \qquad $b\ >\ 0\,$, \qquad $d\ =\ 0$.
\end{description}
We also note that global well-posedness results can be obtained for those multi-symplectic and symplectic Systems~\eqref{eq:5s}, \eqref{eq:6s} with $a\ =\ c\,$, $b\ =\ d$ and satisfying \eqref{eq:const3}.


\subsection{Existence and classification of solitary waves}

A second property to discuss here is the existence of solitary wave solutions
\begin{equation}\label{eq:sw1}
 \zeta\,(x,\,t)\ \eqdef\ \zeta_{\,s}\,(x\ -\ c_{\,s}\,t)\,, \qquad
 u\,(x,\,t)\ \eqdef\ u_{\,s}\,(x\ -\ c_{\,s}\,t)\,,
\end{equation}
where $c_{\,s}$ is the speed of propagation of the waves and the profiles $\eta_{\,s}\ =\ \eta_{\,s}\,(X)\,$, $u_{\,s}\ =\ u_{\,s}\,(X)\,$, $X\ =\ x\ -\ c_{\,s}\,t\,$, are smooth, positive and even functions. Substituting \eqref{eq:sw1} into \eqref{eq:5s} and \eqref{eq:6s} (with $a\ =\ c$), integrating once and setting the integration constant equals zero, we are interested in the profiles solutions of the nonlinear ordinary differential system
\begin{align}\label{eq:sw2}
 -\,c_{\,s}\,\zeta_{\,s}\ +\ u_{\,s}\ +\ \A\,(\zeta_{\,s}\,,\,u_{\,s})\ +\ a\,u_{\,s}^{\,\prime\prime}\ +\ b\,c_{s}\,\zeta_{\,s}^{\,\prime\prime}\ &=\ 0\,, \\
  -\,c_{\,s}\,u_{\,s}\ +\ \zeta_{\,s}\ +\ \B\,(\zeta_{\,s}\,,\,u_{\,s})\ +\ a\,\zeta_{\,s}^{\,\prime\prime}\ +\ d\,c_{s}\, u_{\,s}^{\,\prime\prime} \ &=\ 0\,, \label{eq:sw3}
\end{align}
with $\A\,$, $\B$ satisfying the constraints~\eqref{eq:const}. As in the previous section, our study will be based on the existing results on this subject in the literature for other \textsc{Boussinesq} systems, especially for \eqref{eq:0}, \eqref{eq:00}. In order to derive the corresponding existence results for \eqref{eq:sw2}, \eqref{eq:sw3}, we will focus on the application of the \textit{Normal Form Theory}, (see \eg \cite{Champneys} and references therein). We refer to \cite{DMII} for more details. The application of other existing theories, such as the \textit{Positive Operator Theory}, \cite{BBB}, the \textit{Concentration-Compactness Theory}, \cite{Lions1984, Lions1984a} or \textsc{Toland}'s Theory, \cite{Toland}, will be discussed in Section~\ref{sec:sec4}.

Normal Form Theory is useful to obtain the existence of solutions of \eqref{eq:sw2}, \eqref{eq:sw3} for speeds $c_{\,s}$ greater than but close to one and allows for distinguishing different kinds of solutions depending on the parameters $a\ (\ =\ c)\,$, $b$ and $d\,$. The arguments exposed in, \eg \cite{DMII} for the Systems~\eqref{eq:0}, \eqref{eq:00} can be applied here from rewriting \eqref{eq:sw2}, \eqref{eq:sw3} as a first-order differential system for $U\ =\ {}^{\top}\,(u_{\,1}\,,\,u_{\,2}\,,\,u_{\,3}\,,\,u_{\,4})\ =\ {}^{\top}\,(\zeta_{\,s}\,,\,\zeta_{\,s}^{\,\prime}\,,\,u_{\,s}\,,\,u_{\,s}^{\,\prime}\,)$ as
\begin{align}
  U^{\,\prime}\ &=\ \T\,(U,\,c_{\,s})\ = \ \Lin\,(c_{\,s})\,U\ +\ \Rn\,(U,\,c_{\,s})\,,\label{eq:sw4aaa}\\
  \Lin\,(c_{\,s})\ &\eqdef\ \begin{pmatrix}[1.1]
    0 & 1 & 0 & 0 \\
    \dfrac{d\,c_{\,s}^{\,2}\ +\ a}{\D} & 0 & -\,\dfrac{c_{\,s}}{\D}\;\left(a\ +\ d\right) & 0 \\
    0 & 0 & 0 & 1 \\
    -\,\dfrac{c_{\,s}\,(a\ +\ b)}{\D}& 0 &\dfrac{1}{\D}\;\left(b\,c_{\,s}^{\,2}\ +\ a\right) & 0
   \end{pmatrix}\,, \label{eq:sw4aab} \\
  \Rn\,(U,\,c_{\,s})\ &\eqdef\ \begin{pmatrix}[1.1]
   0 \\
   \dfrac{1}{\D}\;\bigl(-\,d\,c_{\,s}\,\A\,(u_{\,1},\,u_{\,3})\ +\ a\,\B\,(u_{\,1},\,u_{\,3})\bigr) \\
   0 \\
   \dfrac{1}{\D}\;\bigl(-\,b\,c_{\,s}\,\B\,(u_{\,1},\,u_{\,3})\ +\ a\,\A\,(u_{\,1},\,u_{\,3})\bigr)
   \end{pmatrix}\,, \label{eq:sw4aac}
\end{align}
where $\D\ \eqdef\ b\,d\,c_{\,s}^{\,2}\ -\ a^{\,2}$ is assumed to be nonzero. As in \cite{DMII}, the form of the nonlinear term \eqref{eq:sw4aac} makes possible to study the existence of solutions of \eqref{eq:sw4aaa} for positive and small $c_{\,s}\ -\ 1$ by using the linear part \eqref{eq:sw4aab} and the \textit{Normal Form Theory}, see \eg \cite{Champneys} and references therein for details. This leads to the classification displayed in Table~\ref{tab:sw}, where, depending on the choice of the parameters $(a\,$, $b$ and $d)\,$, two types of solitary wave solutions are distinguished: classical solitary waves ({\bfseries\sffamily Class}), that is, smooth and monotonically decaying at infinity travelling wave solutions (or, equivalently, orbits homoclinic to zero at infinity) and generalized solitary waves ({\bfseries\sffamily Gen}) or travelling waves which are homoclinic to periodic solutions at infinity, \cite{Lombardi2000}.

In order to illustrate the way to identify the structure of the homoclinic solutions to \eqref{eq:sw4aaa}, two cases (one for each class of solitary waves) will be described in more detail. The first one takes $a\ =\ c\ =\ 0\,$, $b\ =\ d\ >\ 0\,$. Note first that the system is reversible in the sense that if $\Ss$ is the $4\,\times\,4$ diagonal matrix with diagonal entries given by the vector ${}^{\top}\,(1\,,\,-1\,,\,1\,,\,-1)\,$, then
\begin{equation}\label{eq:rev}
  \Ss\,\T\,(U,\,c_{\,s})\ =\ -\,\T\,(\Ss\,U,\,c_{\,s})\,.
\end{equation}
On the other hand, when $c_{\,s}\ >\ 1\,$, the spectrum of the linearization at the origin $U\ =\ 0$ of \eqref{eq:sw4aaa} consists of four different, real eigenvalues $\bigl\{\,\pm\,\lambda_{\,-},\,\pm\,\lambda_{\,+}\,\bigr\}$ with
\begin{equation*}
  \lambda_{\,\pm}\ =\ \sqrt{\frac{c_{\,s}\ \pm\ 1}{c_{\,s}\,b}}\,,
\end{equation*}
satisfying $\lambda_{\,-}\ <\ 0\ <\ \lambda_{\,+}\,$. When $c_{\,s}\ =\ 1\,$, $\Lin\,(1)$ has two simple eigenvalues $\pm\,\sqrt{2\,/\,b}$ and the zero eigenvalue with geometric multiplicity equals one and algebraic multiplicity equals two. Let $\bigl\{\,w_{\,1}\,,\,w_{\,2}\,,\,w_{\,3}\,,\,w_{\,4}\,\bigr\}$ a basis of generalized eigenvectors, with $w_{\,3}\,$, $w_{\,4}$ eigenvectors of $\sqrt{2\,/\,b}$ and $-\,\sqrt{2\,/\,b}\,$, respectively, $w_{\,1}$ eigenvector associated to the zero eigenvalue and $w_{\,2}$ such that $\Lin\,(1)\,w_{\,2}\ =\ w_{\,1}\,$. Explicitly, we take
\begin{align*}
  w_{\,1}\ =\ {}^{\top}\,(1\,,\,0\,,\,1\,,\,0)\,, \qquad & w_{\,2}\ =\ {}^{\top}\,\biggl(0\,,\,-1\,,\,0\,,\,-1\biggr)\,,\\
  w_{\,3}\ =\ {}^{\top}\,\biggl(\sqrt{\frac{b}{2}}\,,\,1\,,\,-\sqrt{\frac{b}{2}}\,,\,-1\biggr)\,, \qquad & w_{\,4}\ =\ {}^{\top}\,\biggl(\sqrt{\frac{b}{2}}\,,\,1\,,\,-\sqrt{\frac{b}{2}}\,,\,-1\biggr)\,.
\end{align*}
Note that $w_{\,1}\,$, $w_{\,2}$ additionally satisfy $\Ss\,w_{\,1}\ =\ w_{1}\,$, $\Ss\,w_{\,2}\ =\ -\,w_{\,2}\,$. If $\Pp$ is the $4\,\times\,4$ matrix with columns given by the $w_{\,j}\,$'s, then we consider the new variables $V\ =\ {}^{\top}\,(v_{\,1}\,,\,v_{\,2}\,,\,v_{\,3}\,,\,v_{\,4})$ such that $U\ =\ \Pp\,V\,$. If $\mu\ \eqdef\ c_{\,s}\ -\ 1\ >\ 0\,$, the System~\eqref{eq:sw4aaa} in the new variables reads
\begin{eqnarray}\label{eq:ns1}
  v_{\,1}^{\,\prime}\ &=&\ v_{\,2}\,, \\
  v_{\,2}^{\,\prime}\ &=&\ -\frac{\mu\,v_{\,1}}{b}\ +\ \frac{v_{\,1}^{\,2}}{b}\;\sum_{i,\,j\ =\ 1}^{2}(\alpha_{\,i\,j}\ +\ \beta_{\,i\,j})\ +\ \O\,\bigl(\,\mu\,\abs{V}\ +\ \abs{V}^{\,2}\,\bigr)\,,\label{eq:ns2} \\
  v_{\,3}^{\,\prime}\ &=&\ \sqrt{\frac{2}{b}}\;v_{\,3}\ +\ \O\,\bigl(\,\mu\,\abs{V}\ +\ \abs{V}^{\,2}\,\bigr)\,,\label{eq:ns3} \\
  v_{\,4}^{\,\prime}\ &=&\ -\,\sqrt{\frac{2}{b}}\;v_{\,4}\ +\ \O\,\bigl(\,\mu\,\abs{V}\ +\ \abs{V}^{\,2}\,\bigr)\,,\label{eq:ns4}
\end{eqnarray}
as $\mu\,$, $\abs{V}\ \rightarrow\ 0$ ($\abs{V}$ stands for the \textsc{Euclidean} norm of $V$ in $\R^{\,4}$) and where we assume that
\begin{equation}
  \sum_{i,\,j\ =\ 1}^{2}(\alpha_{\,i\,j}\ +\ \beta_{\,i\,j})\ >\ 0\,. \label{eq:cond}
\end{equation}
Then, for $\mu$ positive and sufficiently small, the center-manifold reduction theorem, \cite{IoosA}, can be applied to see that bounded solutions of the System \eqref{eq:ns1} -- \eqref{eq:ns4} are on a locally invariant, center manifold which determines a dependence $(v_{\,3},\,v_{\,4})\ =\ h\,(\,\mu,\,v_{\,1},\,v_{\,2}\,)$ for some smooth $h\,(\,\mu,\,v_{\,1},\,v_{\,2}\,)\ =\ \O\,\bigl(\mu\,\abs{{}^{\top}\,(v_{\,1},\,v_{\,2})}\ +\ \abs{{}^{\top}\,(v_{\,1},\,v_{\,2})}^{\,2}\bigr)$ as $\mu\,$, $\abs{{}^{\top}\,(v_{\,1},\,v_{\,2})}\ \rightarrow\ 0\,$, see \cite[Theorem 3.2]{IoosK}. Furthermore, every solution $v_{\,1}\,$, $v_{\,2}$ of the reduced system obtained from \eqref{eq:ns1}, \eqref{eq:ns2} with $(v_{\,3},\,v_{\,4})\ =\ h\,(\,\mu,\,v_{\,1},\,v_{\,2}\,)$ leads to a solution of Equations~\eqref{eq:ns1} -- \eqref{eq:ns4} through this dependence. The normal form system can be written as
\begin{equation}
  v_{\,1}^{\,\prime}\ =\ v_{\,2}\,,\qquad v_{\,2}^{\,\prime}\ =\ c_{\,1}\,(\mu)\,v_{\,1}\ +\ c_{\,2}\,(\mu)\,v_{\,2}\,,\label{eq:nf}
\end{equation}
for some coefficients $c_{\,1}\,(\mu)\,$, $c_{\,2}\,(\mu)\,$. A suitable $\mu-$scaling of the variables transforms \eqref{eq:nf} into
\begin{equation}\label{eq:snf}
  v_{\,1}^{\,\prime}\ =\ v_{\,2}\,,\quad v_{\,2}^{\,\prime}\ =\ \sign\,(\mu)\,v_{\,1}\ -\ \frac{3}{2}\;v_{\,1}^{\,2}\ +\ \O\,(\mu)\,,
\end{equation}
For $\mu\ >\ 0\,$, \eqref{eq:snf} admits a solution of the form $v_{\,1}\,(x)\ =\ \sech^{\,2}\,(x\,/\,2)\ +\ \O\,(\mu)\,$, $v_{\,2}\ =\ v_{\,1}^{\,\prime}\,$. The persistence of this homoclinic orbit from the perturbation which connects to the original System \eqref{eq:sw4aaa} can be proved by using similar arguments to those of \cite{IoosK} (see also \cite{Champneys}).

\begin{figure}
  \centering
  \subfigure[]
  {\includegraphics[width=0.49\textwidth]{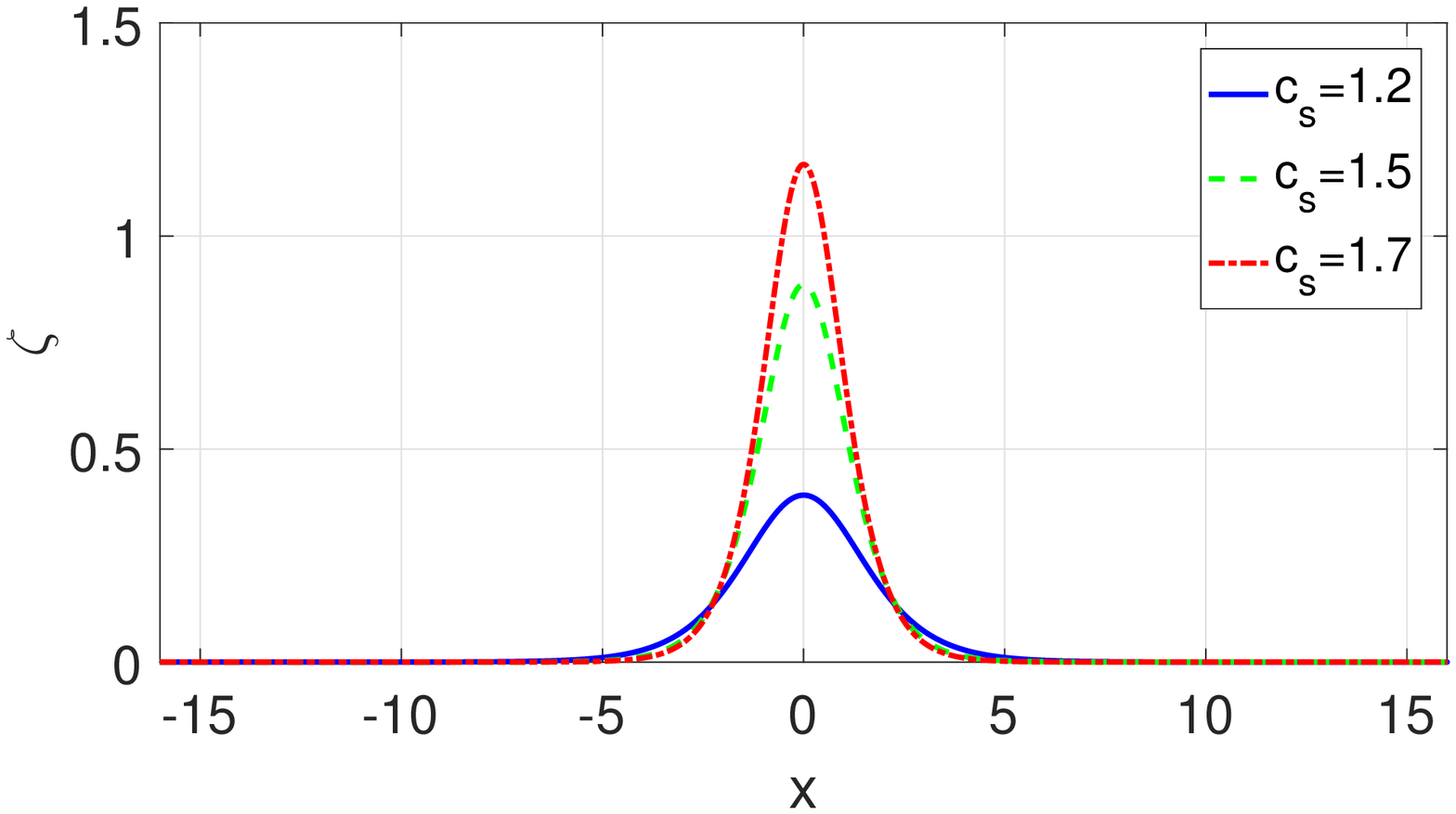}}
  \subfigure[]
  {\includegraphics[width=0.49\textwidth]{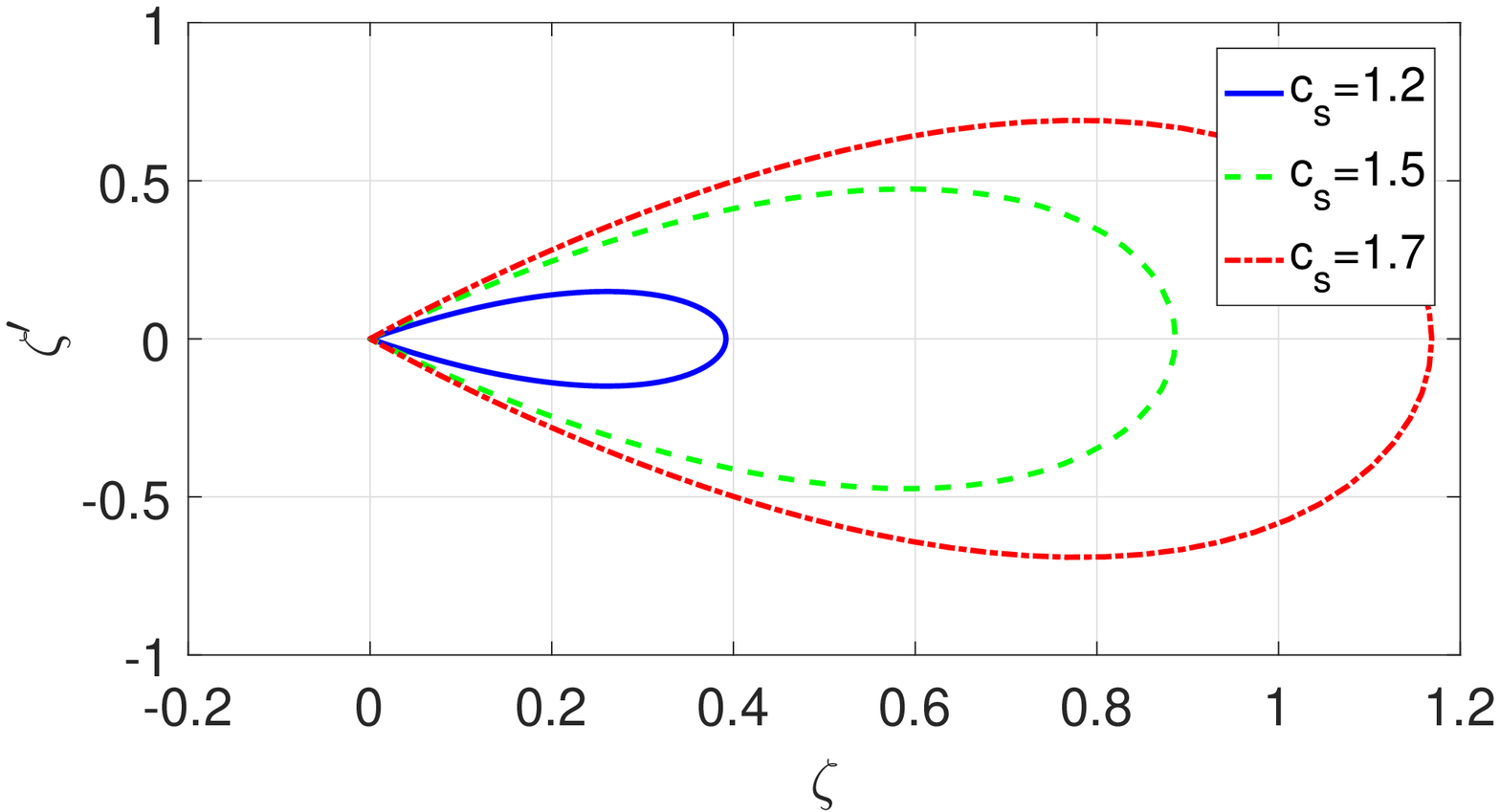}}
  \subfigure[]
  {\includegraphics[width=0.49\textwidth]{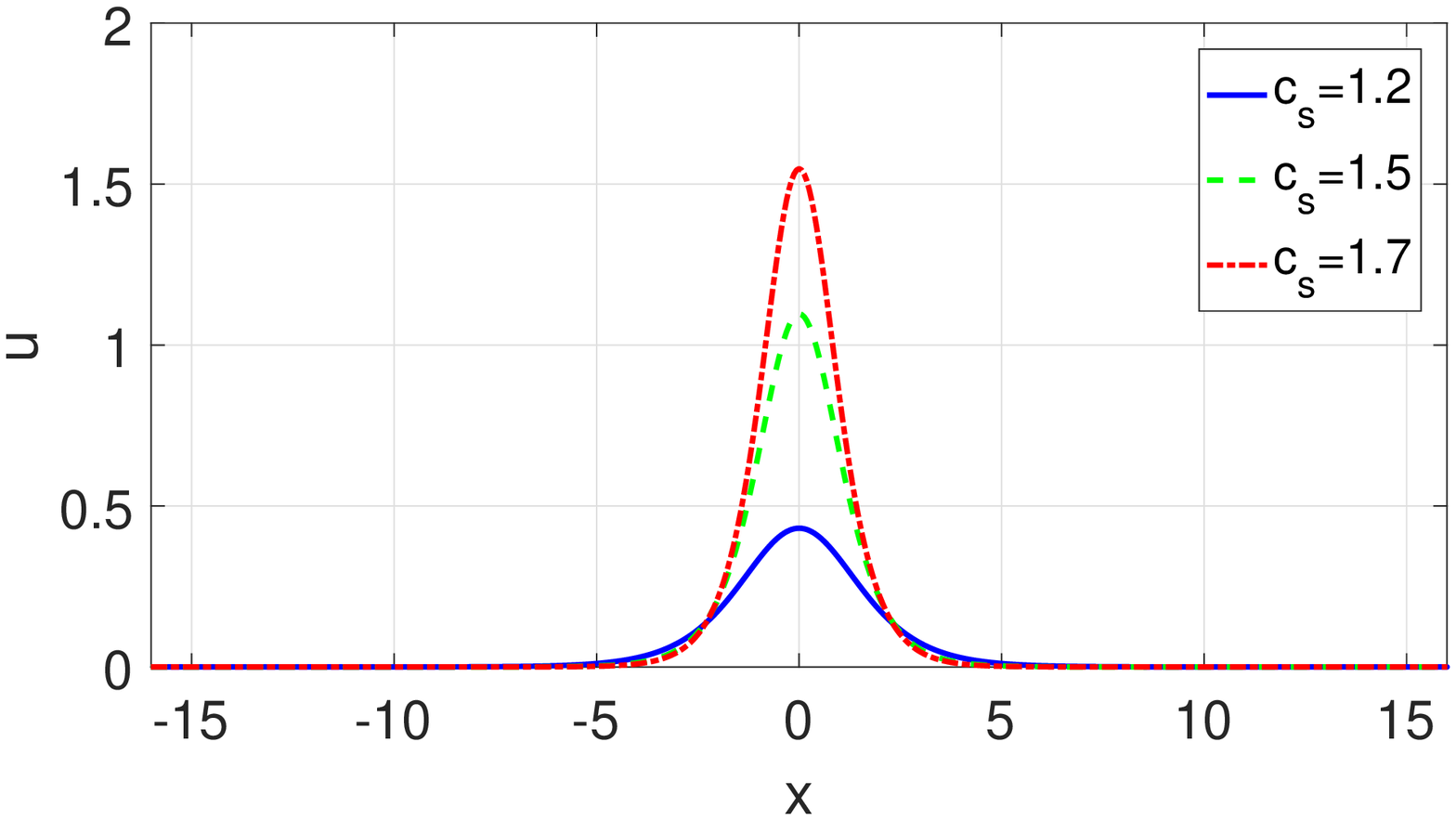}}
  \subfigure[]
  {\includegraphics[width=0.49\textwidth]{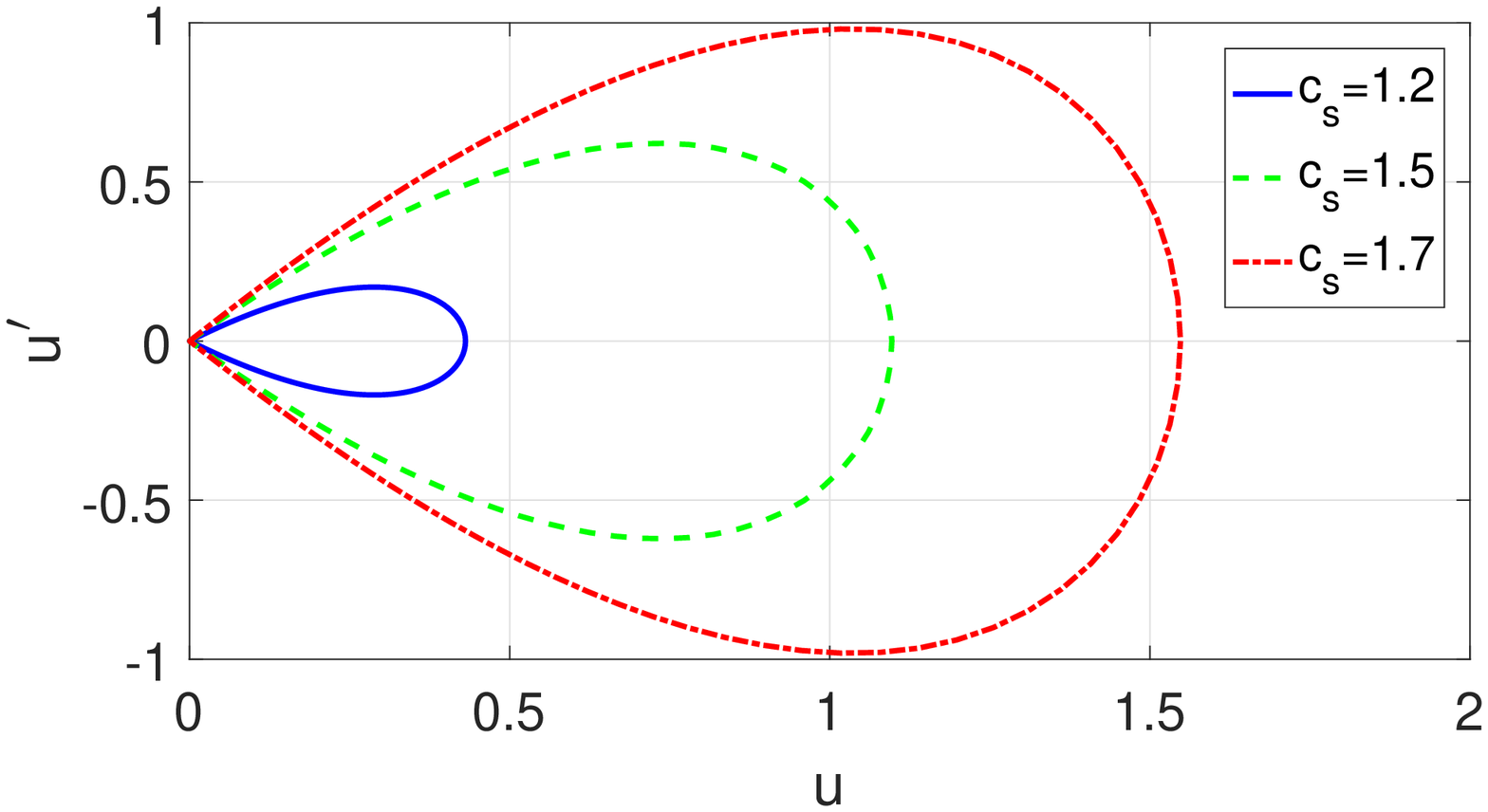}}
  \caption{\small\em Approximate classical solitary wave profile solutions of Equations~\eqref{eq:sw2} with $\alpha_{\,1\,1}\ =\ 0\,$, $\alpha_{\,1\,2}\ =\ 0.46\,$, $\alpha_{\,2\,2}\ =\ 0\,$, $\beta_{\,1\,1}\ =\ 0.23\,$, $\beta_{\,1\,2}\ =\ 0\,$, $\beta_{\,2\,2}\ =\ 0.73$ and $a\ =\ c\ =\ 0\,$, $b\ =\ d\ =\ 1/6\,$. (a) $\zeta$ profiles; (b) phase portrait of (a); (c) $u$ profiles; (d) phase portrait of (c).}
  \label{Fig:swp1}
\end{figure}

The form of the waves is illustrated in Figures~\ref{Fig:swp1}, which displays $\zeta$ and $u$ profiles (with the corresponding phase portraits) solutions of \eqref{eq:sw2}, \eqref{eq:sw3}, with the coefficients specified in the caption, and for several values of the speed $c_{\,s}\,$. 

\begin{table}
  \centering
  \caption{\small\em Classification of solitary wave solutions using normal form theory for small and positive values of $c_{\,s}\ -\ 1\,$.}
  \bigskip
  \begin{tabular}{c|c}
  \hline\hline
  \textit{Parameters} & \textit{Solitary wave type} \\
  \hline\hline
  $a\ =\ c\ >\ 0\,$, $b\ =\ d\ =\ 0$ & {\bfseries\sffamily Gen} \\
  $a\ =\ c\ <\ 0\,$, $b\ =\ 0\,$, $d\ >\ 0$ & {\bfseries\sffamily Gen} \\
  $a\ =\ c\ >\ 0\,$, $d\ =\ 0\,$, $b\ >\ 0$ & {\bfseries\sffamily Gen} \\
  $a\ =\ c\ <\ 0\,$, $b\ >\ 0\,$, $d\ >\ 0\,$, $b\,d\ -\ a^{\,2}\ >\ 0$ & {\bfseries\sffamily Class} \\
  $a\ =\ c\ <\ 0\,$, $b\ >\ 0\,$, $d\ >\ 0\,$, $b\,d\ -\ a^{\,2}\ <\ 0$ & {\bfseries\sffamily Gen} \\
  $a\ =\ c\ >\ 0\,$, $b\ >\ 0\,$, $d\ >\ 0\,$, $b\,d\ -\ a^{\,2}\ >\ 0$ & {\bfseries\sffamily Class} \\
  $a\ =\ c\ >\ 0\,$, $b\ >\ 0\,$, $d\ >\ 0\,$, $b\,d\ -\ a^{\,2}\ <\ 0$ & {\bfseries\sffamily Gen} \\
  $a\ =\ c\ >\ 0\,$, $b\ =\ d\ <\ 0$ & {\bfseries\sffamily Gen} \\
  $a\ =\ c\ =\ 0\,$, $b\ >\ 0\,$, $d\ >\ 0\,$& {\bfseries\sffamily Class}\smallskip \\
  \hline\hline
  \end{tabular}
  \label{tab:sw}
\end{table}

\begin{figure}
  \centering
  \subfigure[]{\includegraphics[width=0.79\textwidth]{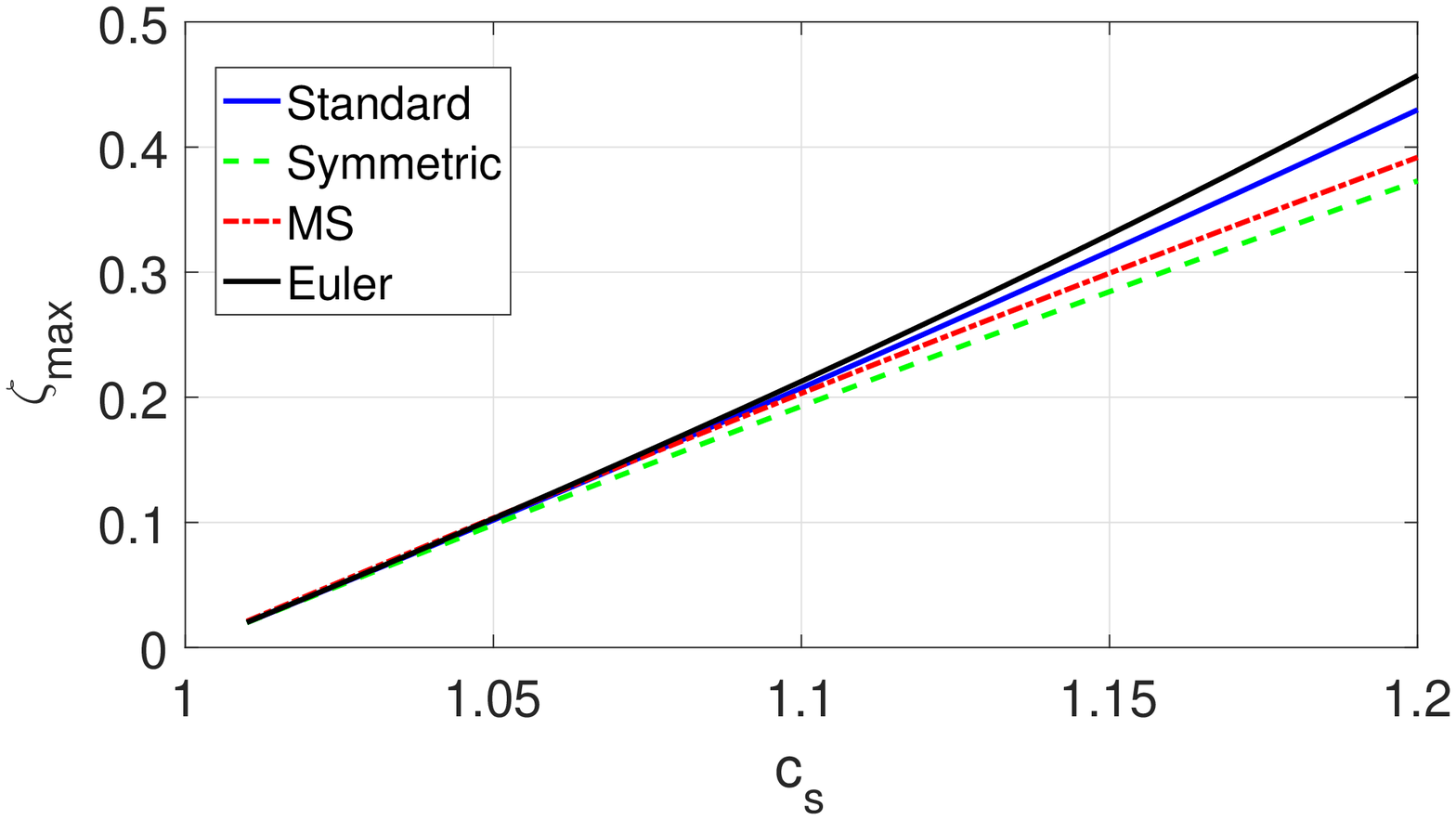}\bigskip}
  \bigskip
  \subfigure[]{\includegraphics[width=0.79\textwidth]{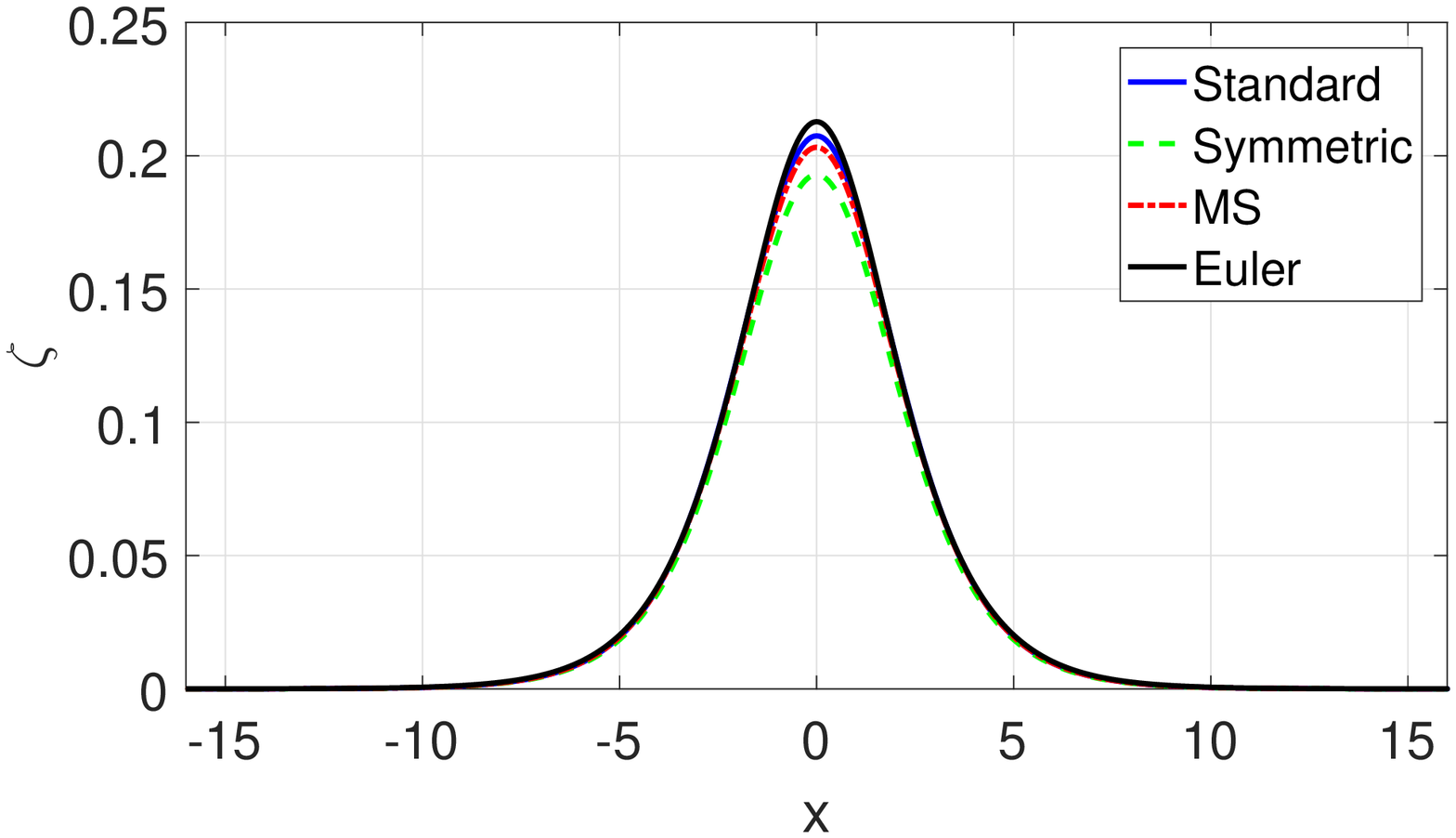}}
  \caption{\small\em (a) Speed-amplitude relation; (b) Comparison of $\zeta$ profiles for $c_{\,s}\ =\ 1.1\,$.}
  \label{Fig:swp2}
\end{figure}

\begin{remark}
As mentioned in the Introduction (see also Section~\ref{sec:sec4}), the question of the consistency of the \textsc{Euler} equations with the $(a\,,\,b\,,\,a\,,\,b)$ MS Systems~\eqref{eq:5s}, \eqref{eq:6s} will be a subject of future research. Some related preliminary remarks can be made here. Recall first that the consistency of the \textsc{Euler} equations with the Systems~\eqref{eq:1}, \eqref{eq:2} was established in \cite{BCL}. (For the case of \eqref{eq:0}, \eqref{eq:00}, see \cite{Bona2004}.) In particular, consistency holds for the MS family ($a\ =\ c$). On the other hand, Figure~\ref{Fig:swp2} compares the solitary-wave speed-amplitude relations, for the case $a\ =\ c\ =\ 0\,$, $b\ =\ d\ >\ 0\,$, of the Systems~\eqref{eq:0}, \eqref{eq:00}, the symmetric version \eqref{eq:1}, \eqref{eq:2} and the MS System~\eqref{eq:5s}, \eqref{eq:6s} used in Figure~\ref{Fig:swp1} with those of the \textsc{Euler} system, computed as in \cite{Clamond2012b, Dutykh2013b}. These results (and those of other experiments not shown here) suggest the existence of MS Systems~\eqref{eq:5s}, \eqref{eq:6s}, close to the symmetric \eqref{eq:1}, \eqref{eq:2}, with a comparable speed-amplitude behaviour.
\end{remark}

In order to illustrate the arguments leading to the existence of generalized solitary waves of Table~\ref{tab:sw}, we will consider the case $a\ =\ c\ >\ 0\,$, $b\ =\ d\ =\ 0$ (see \cite{Bona2007} for the KdV--KdV System \eqref{eq:0}, \eqref{eq:00}). Note first that \eqref{eq:sw4aaa} is still reversible as it satisfies \eqref{eq:rev}. Now, as for the spectrum of the matrix $\Lin$ in \eqref{eq:sw4aab} at $c_{\,s}\ =\ 1\,$, the difference with the case of classical solitary waves, considered above, is the presence of two simple, pure imaginary eigenvalues $\pm\,\ui\,\sqrt{2\,/\,a}\,$, while zero is also an eigenvalue with geometric multiplicity one and algebraic multiplicity two. The basis $\bigl\{w_{\,1},\,w_{\,2},\,w_{\,3},\,w_{\,4}\,\bigr\}\,$, in $\mathds{C}^{\,4}\,$, with $w_{\,1}\,$, $w_{\,2}$ as above, contains the eigenvectors
\begin{align*}
  w_{\,3}\ &=\ {}^{\top}\,\biggl(\,\ui\,\sqrt{\frac{a}{2}}\,,\,1\,,\,-\ui\,\sqrt{\frac{a}{2}}\,,\,-1\biggr)\,,\\ 
  w_{\,4}\ &=\ {}^{\top}\,\biggl(\ui\,\sqrt{\frac{a}{2}}\,,\,1\,,\,-\ui\,\sqrt{\frac{a}{2}}\,,\,-1\biggr)\,,
\end{align*} 
associated to $\ui\,\sqrt{2\,/\,a}$ and $-\ui\,\sqrt{2\,/\,a}\,$, respectively. The application of the \textit{Normal Form Theory} in this case may follow the direct approach of \cite[Section~32]{IoosK} or, alternatively, that of \textsc{Lombardi} in \cite{Lombardi2000}. In this last case, let $\bigl\{\,w_{\,1}^{\,*}\,,\,w_{\,2}^{\,*}\,,\,w_{\,3}^{\,*}\,,\,w_{\,4}^{\,*}\bigr\}$ be the corresponding dual basis (with, in particular, $w_{\,1}^{\,*}\ =\ {}^{\top}\,(0\,,\,1/2\,,\,0\,,\,1/2)$). If $\DD_{\,\mu,\,U}^{\,2}\;\T\,(U,\,\mu)$ denotes the derivative, with respect to $\mu\,$, of the \textsc{Jacobian} matrix of $\T$ in \eqref{eq:sw4aaa} and $\DD_{\,U,\,U}\;\T\,(U,\,\mu)^{\,2}$ the \textsc{Hessian} (matrix) of $\T\,$, then
\begin{align*}
  c_{\,1\,0}\ &\eqdef\ \langle\, w_{\,1}^{\,*},\, \DD_{\,\mu,\,U}^{\,2}\;\T\,(0,\,0)\,w_{\,0}\,\rangle\ =\ \frac{1}{a}\ >\ 0\,, \\
  c_{\,2\,0}\ &\eqdef\ \frac{1}{2}\;\langle\,w_{\,1}^{\,*},\,\DD_{\,U,\,U}^{\,2}\;\T\,(0,\,0)\,[\,w_{\,0},\,w_{\,0}\,]\,\rangle\ =\ -\,\frac{1}{2\,a}\;\sum_{i,\,j\ =\ 1}^{\,2}\,(\alpha_{\,i\,j}\ +\ \beta_{\,i\,j})\ \neq\ 0\,,
\end{align*}
where \eqref{eq:cond} is assumed. Therefore \cite[Theorem~7.1.1]{Lombardi2000} applies to prove that, for $\mu$ small enough, $\T\,(U,\,\mu)$ admits, near the fixed point $U\ =\ 0\,$, a one-parameter family of periodic orbits of arbitrarily small amplitude (depending on $\mu$) and a pair of reversible, homoclinic connections to them. A direct application of \textit{Normal Form Theory} to the reduced system, in a similar way to the case of classical solitary waves, can be seen in \cite{IoosK}.

\begin{figure}
  \centering
  \subfigure[]
  {\includegraphics[width=0.49\textwidth]{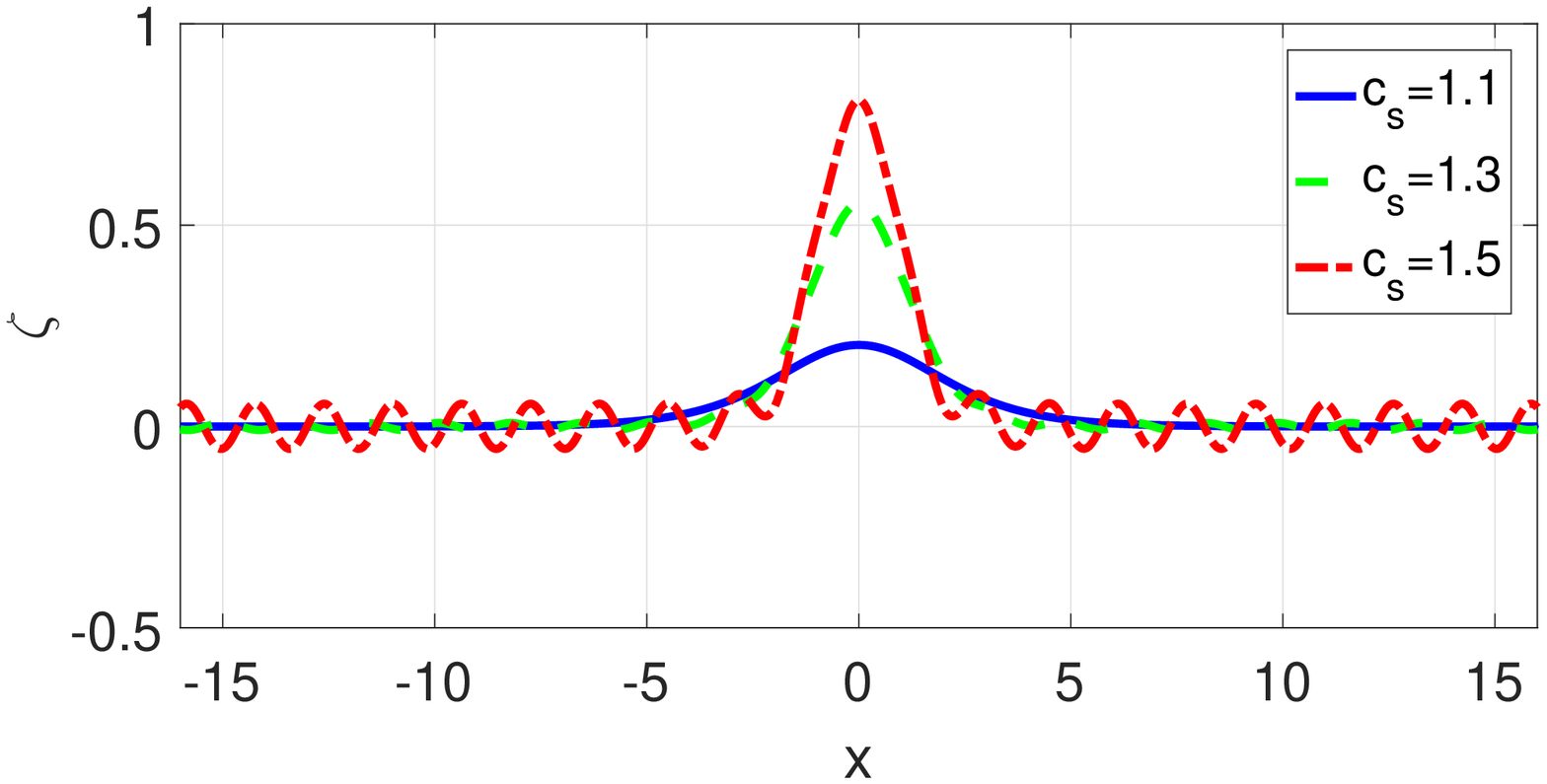}}
  \subfigure[]
  {\includegraphics[width=0.49\textwidth]{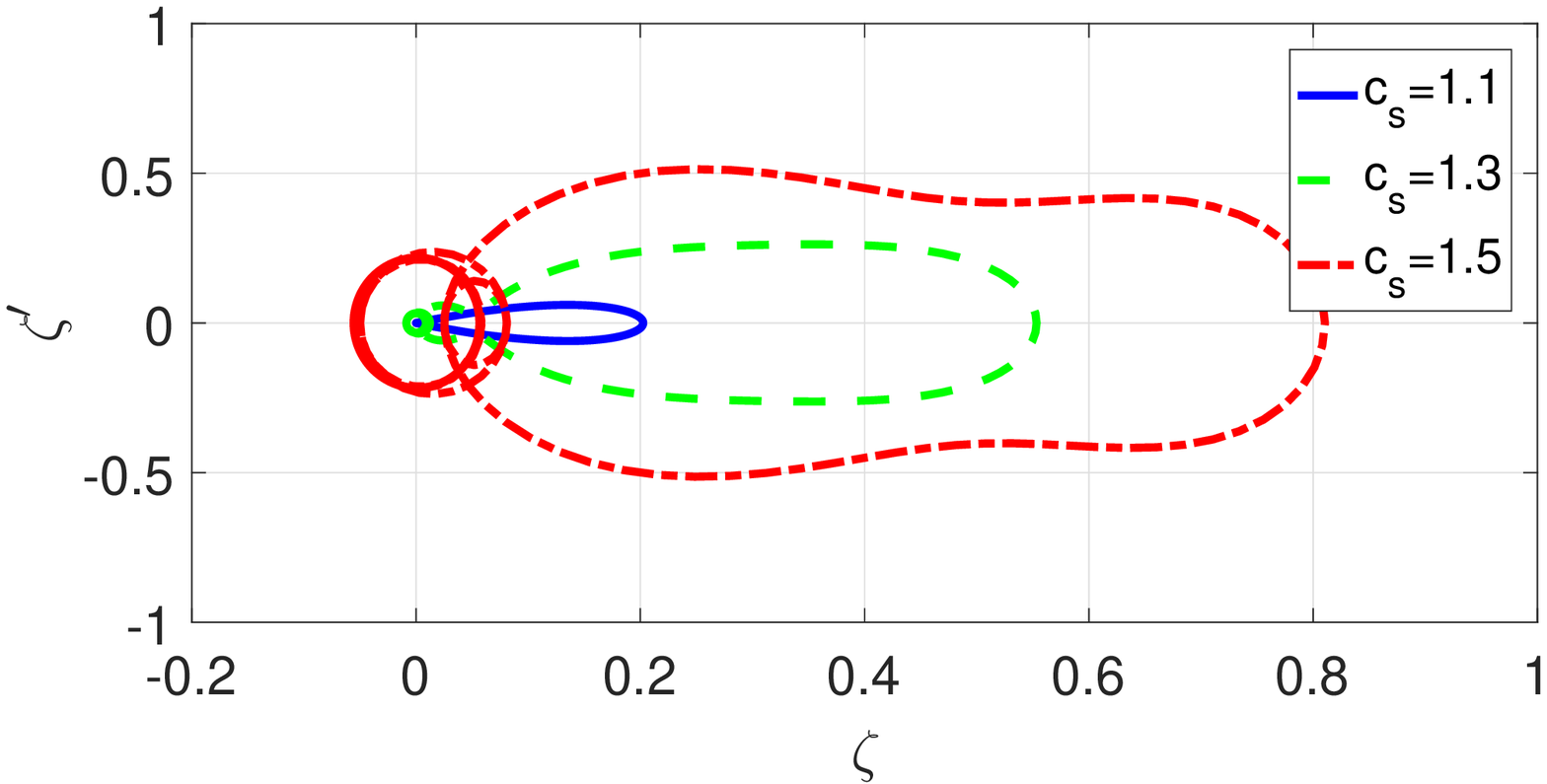}}
  \subfigure[]
  {\includegraphics[width=0.49\textwidth]{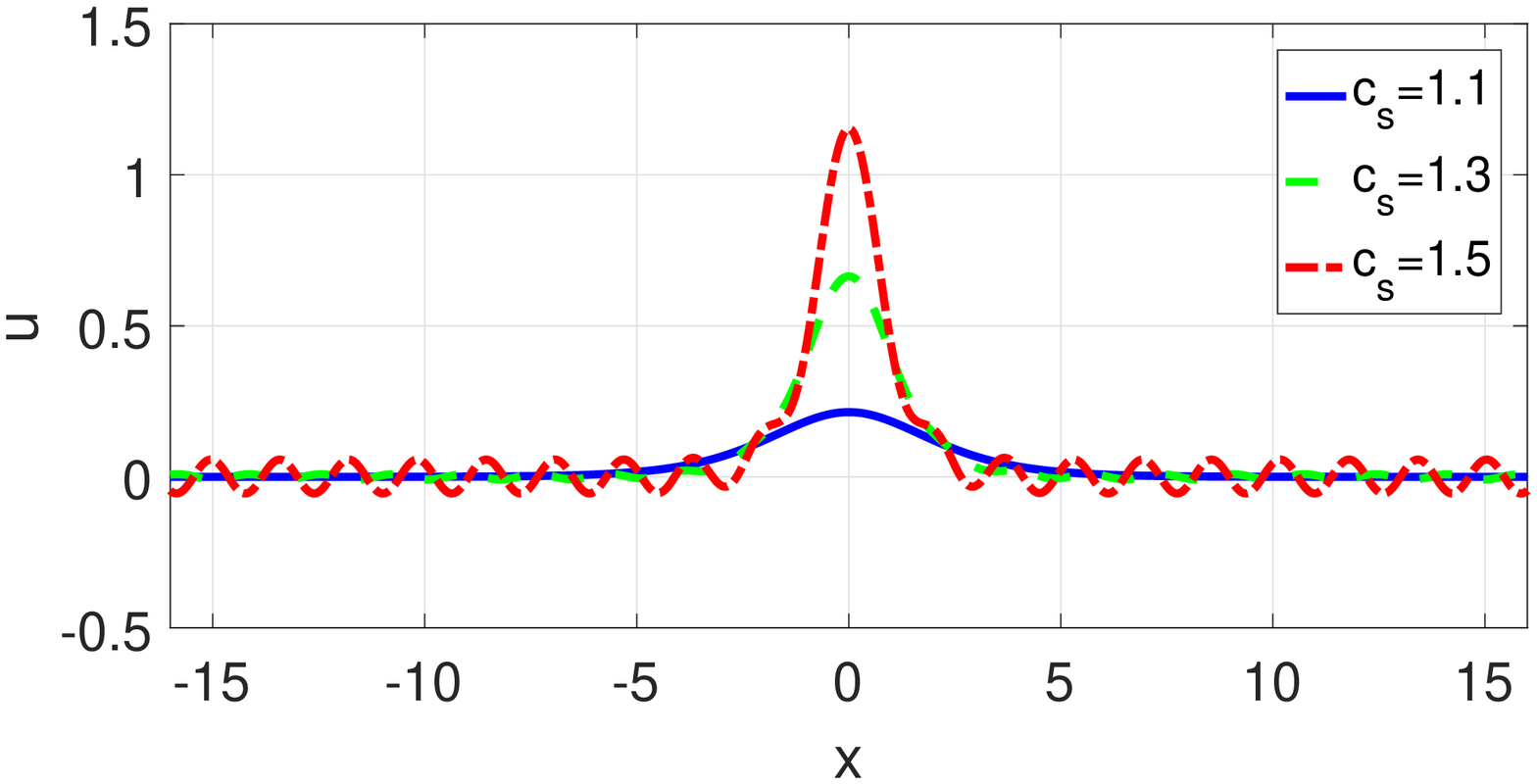}}
  \subfigure[]
  {\includegraphics[width=0.49\textwidth]{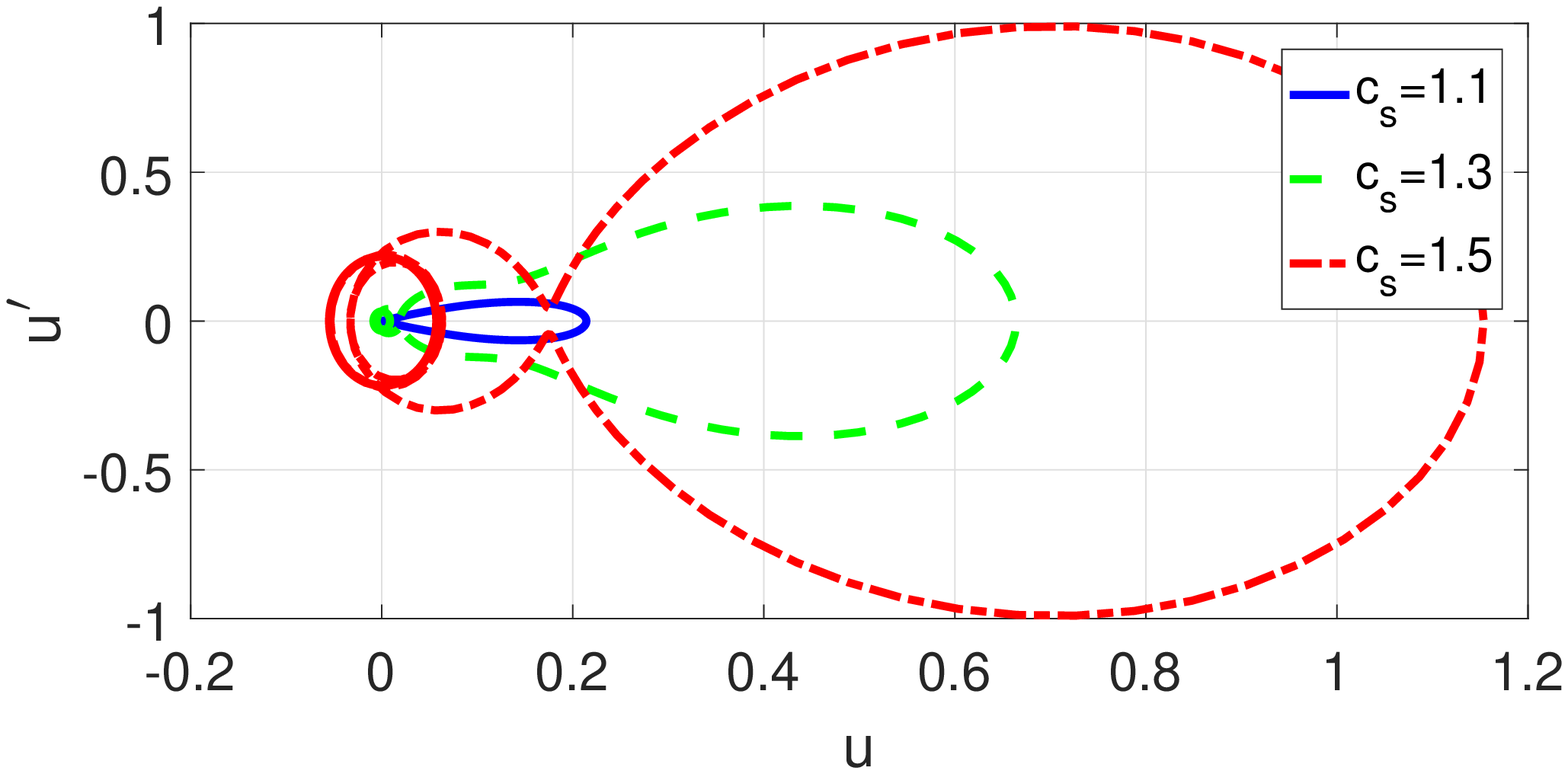}}
  \caption{\small\em Approximate generalized solitary wave profile solutions of Equations~\eqref{eq:sw2} with $\alpha_{\,1\,1}\ =\ 0\,$, $\alpha_{\,1\,2}\ =\ 0.46\,$, $\alpha_{\,2\,2}\ =\ 0\,$, $\beta_{\,1\,1}\ =\ 0.23\,$, $\beta_{\,1\,2}\ =\ 0\,$, $\beta_{\,2\,2}\ =\ 0.73$ and $a\ =\ c\ =\ 1/6\,$, $b\ =\ d\ =\ 0\,$. (a) $\zeta$ profiles; (b) phase portraits of (a); (c) $u$ profiles; (d) phase portraits of (c).}
  \label{Fig:swp3}
\end{figure}

\begin{figure}
  \centering
  \subfigure[]
  {\includegraphics[width=0.49\textwidth]{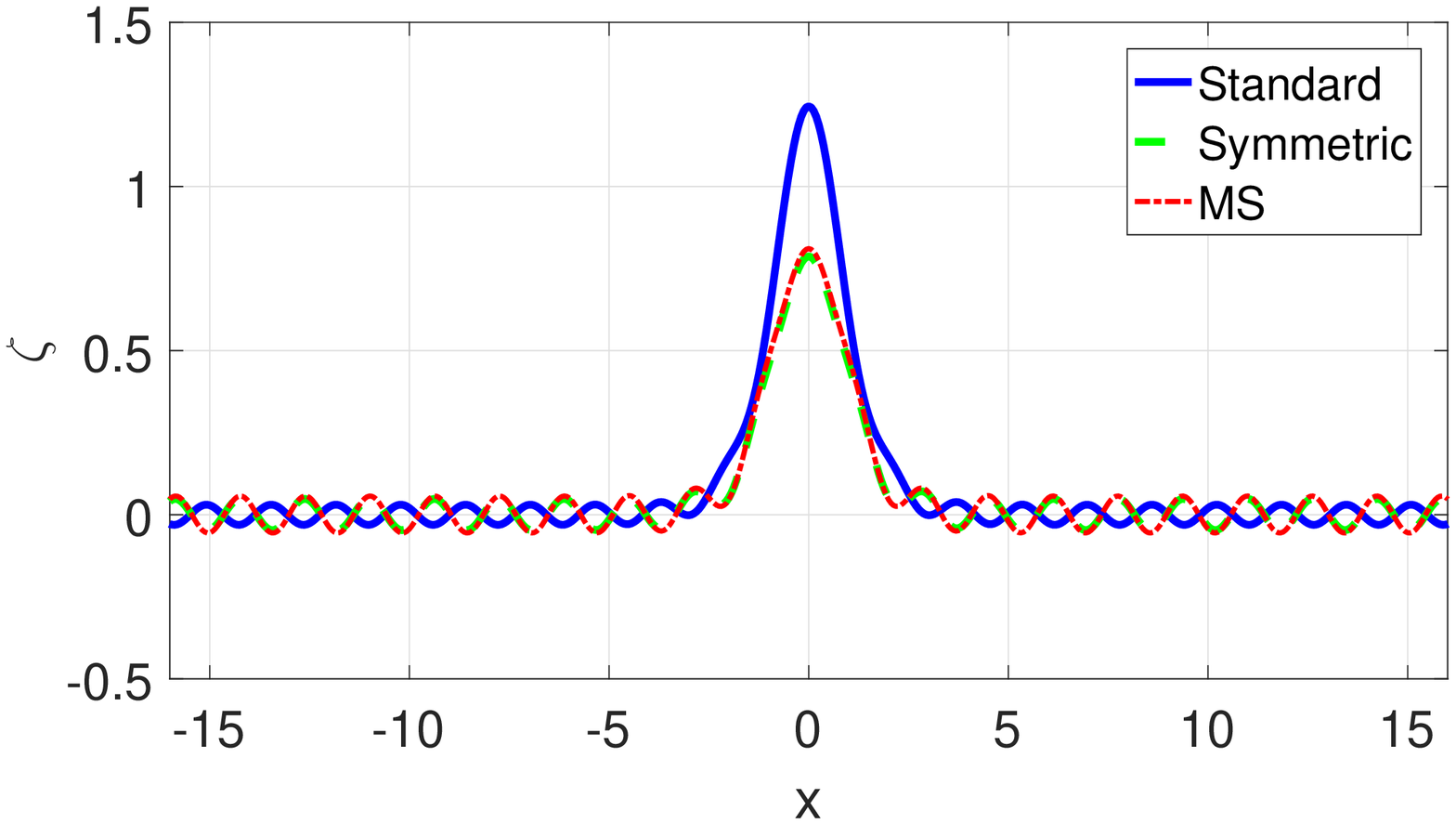}}
  \subfigure[]
  {\includegraphics[width=0.49\textwidth]{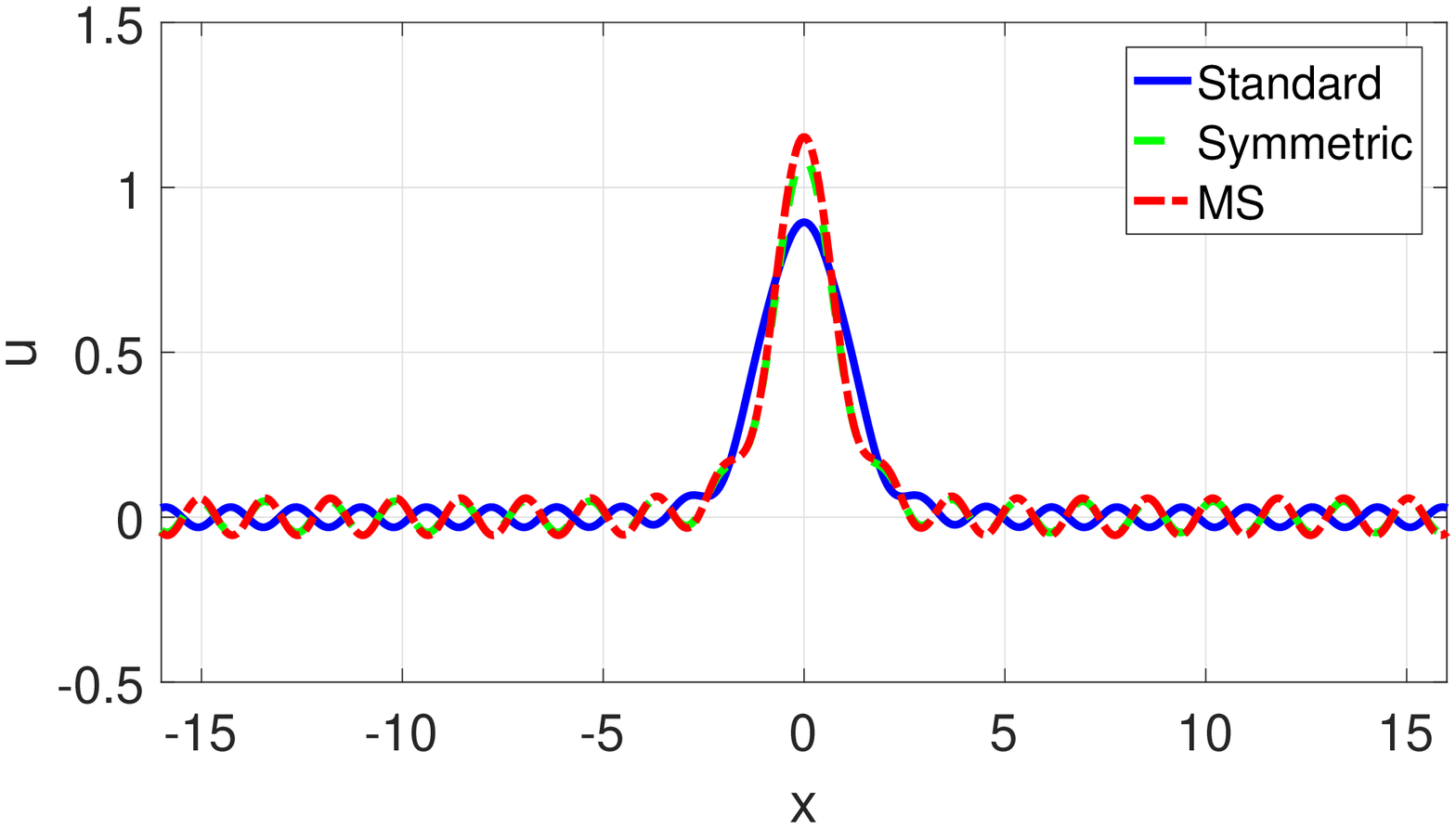}}
  \caption{\small\em Comparison with KdV--KdV System \eqref{eq:0}, \eqref{eq:00} the symmetric version \eqref{eq:1}, \eqref{eq:2} and the MS System \eqref{eq:5s}, \eqref{eq:6s} of Figure~\ref{Fig:swp3} for $c_{\,s}\ =\ 1.5\,$. (a) $\zeta$ profiles; (b) $u$ profiles.}
  \label{Fig:swp4}
\end{figure}

The form of the waves is illustrated in Figure~\ref{Fig:swp3} for different speeds. The coefficients of the nonlinear part of the MS system are the same as those of Figure~\ref{Fig:swp1} while $a\ =\ c\ =\ 1/6\,$, $b\ =\ d\ =\ 0\,$. Finally, Figure~\ref{Fig:swp4} compares the approximate generalized profile corresponding to $c_{\,s}\ =\ 1.5$ with those of the KdV--KdV System \eqref{eq:0}, \eqref{eq:00} and of the symmetric version \eqref{eq:1}, \eqref{eq:2}.

\section{Discussion}
\label{sec:sec4}

Below we discuss the main conclusions and perspectives of our study.


\subsection{Conclusions}

In the present manuscript we extended the class of known multi-symplectic systems arising in the modeling nonlinear shallow water waves, with especial emphasis on \textsc{Boussinesq}-type equations for surface wave propagation. Having in mind that the full \textsc{Euler} equations are multi-symplectic, \cite{Bridges1997}, it seems reasonable to expect from an approximate model to preserve this property. This is the case of, for example, the fully nonlinear SGN equations, \cite{Chhay2016, Clamond2016a}, or the simpler, but fundamental, NLS, KdV and BBM equations, \cite{Chen2002, Zhao2000, Bridges2001, Sun2004}. Because of the last two, it is expected for the KdV--BBM equation to be the next example. The MS structure of the KdV--BBM equation was derived in this paper and with a procedure that can be extended to other equations. Focused on \textsc{Boussinesq} systems, our first result was the proof of the MS structure of the symmetric, four-parameter $(a\,,\,b\,,\,a\,,\,d)$ family proposed in \cite{BCL}. This is, however, a particular case of a second, more general approach. The main steps of the procedure helped in constructing a family of MS \textsc{Boussinesq}-type systems, from a four-parameter $(a\,,\,b\,,\,c\,,\,d)$ structure in the dispersive part and general homogeneous quadratic polynomials as nonlinear terms, by identifying those combinations of the parameters leading to the MS structure. Two main consequences of the method can be emphasized. The first one is that no one of the $(a\,,\,b\,,\,a\,,\,d)$ sub-family of \textsc{Boussinesq} systems, introduced and analyzed in \cite{BCS, Bona2004}, is multi-symplectic; however, a modification of the nonlinear part (which, in scaled variables, belongs to a higher order of the asymptotic derivation and therefore is asymptotically negligible compared to the first-order terms) enables us to obtain a related sub-family of MS \textsc{Boussinesq} equations. The second consequence is the derivation of a four-parameter, symplectic and multi-symplectic class of equations, with two parameters in the dispersive part and two from the nonlinearity. In addition to this, some mathematical properties of the new systems were studied. The discussion was focused on well-posedness of the corresponding IVP and the existence of solitary wave solutions. The derivation of results makes use of the pertinent literature for other \textsc{Boussinesq} systems.


\subsection{Perspectives}

The future research may follow several directions. The first one concerns the consistency of the \textsc{Euler} system with the new family of MS \textsc{Boussinesq}-type equations. The study on this matter developed in \cite{BCL} can be considered as starting point of our future analysis, as well as some comparison of the speed-amplitude relation for classical solitary waves computed in Section~\ref{sec:sec3}. In the first case, the results concern the MS family within the symmetric Systems~\eqref{eq:1}, \eqref{eq:2}. The second point is illustrated here by comparing the speed-amplitude relation for solitary waves of the \textsc{Euler} equations with the Systems~\eqref{eq:0}, \eqref{eq:00} and MS Systems~\eqref{eq:5s}, \eqref{eq:6s}. The results suggest that close to the MS family in \eqref{eq:1}, \eqref{eq:2} there are systems improving this relation and the corresponding approximation to the solitary waves of the \textsc{Euler} system. This may be understood as an indication of consistency and deserves to be explored in more detail in a future research.

On the other hand, the results on the existence of solitary wave solutions derived in this paper made use of the Normal Form Theory, following previous applications, \cite{Champneys, DMII}. This has the limitation of the local character of this theory, leading to an existence of solutions which is ensured for speeds $c_{\,s}$ greater than but close enough to one. For the case of classical solitary waves, some other strategies for the existence are available. Of particular interest for us are two:
\begin{itemize}
  \item The Positive Operator Theory, \cite{BBB}, which is based on writing the equations for the traveling waves in the form of a fixed-point system and analyzing the existence of solution from the properties of the fixed-point operator on cones in a suitable \textsc{Fr\'echet} space. In this sense, we think that the approach carried out by \textsc{Bona} and \textsc{Chen} in \cite{BonaCh} to derive existence results of classical solitary waves of \textsc{Boussinesq}-type systems can be applied to the new MS equations. The corresponding results would not be limited by the magnitude of $c_{\,s}\ -\ 1\,$.
  \item A second, interesting for us, approach is \textsc{Toland}'s Theory, \cite{Toland}. This is based on the qualitative theory for second-order differential systems and, under suitable conditions, determines a curve $f\ =\ 0$ satisfying that the corresponding solution with initial data on it leads to an orbit which is homoclinic to zero at infinity, being therefore identified as a classical solitary wave profile. In order to adapt this theory to the new systems, two first questions should be overcome (\cf \cite{DMII} and references therein). The first one concerns the symmetric, conservative form in which the system for the profiles must be written. The second one is the potential energy associated to the nonlinear part of the resulting equations. Due to the general quadratic form of the nonlinearity of the new systems, this potential energy is different and leads to a curve of initial data of cubic degree, unlike other cases like some $(a\,,\,b\,,\,c\,,\,d)$ systems of \cite{BCS, Bona2004}, where $f$ is quadratic. Finally, the indefinite character of the kinetic energy of the steady system \eqref{eq:sw2}, \eqref{eq:sw3} (leading to the condition $a\cdot c_{\,s}\cdot (b\ -\ d)\ <\ 0$) must be considered.
\end{itemize}

Concerning the stability of the solitary waves, the \textsc{Hamiltonian} structure is known to be used in some cases to analyze the orbital stability, \eg \cite{Weinstein1986, Grillakis1987, Grillakis1990}, using the characterization of the solitary waves as equilibria of the \textsc{Hamiltonian} relative to a fixed value of the momentum. In addition to this, \textsc{Bridges} and \textsc{Derks}, \cite{Bridges1999}, present a characterization of the \textsc{Evans} functions in \textsc{Hamiltonian} PDEs and use it to analyze the linear stability of three examples: the classical semilinear wave equation, the generalized KdV equation and a \textsc{Boussinesq} system. These results suggest as future research the use of the multi-symplectic structure to investigate some aspects of the stability of the solitary waves of the family of MS \textsc{Boussinesq}-type equations.

Finally, our study opens new perspectives in the construction of structure-preserving integrators for \textsc{Boussinesq}-type equations, since only a sub-family $(a,\,b,\,c,\,b)$ possesses the \textsc{Hamiltonian} formulation to the best of our knowledge. Fortunately, all of them are multi-symplectic. Thus, with the multi-symplectic structure unveiled in this paper, the so-called multi-symplectic integrators can be readily applied to \textsc{Boussinesq}-type systems as well. We remind that these discretizations can be of finite difference \cite{Bridges2001} or of pseudo-spectral \cite{Islas2003} types to satisfy the most stringent accuracy requirements. Moreover, the major advantage of these methods is that they preserve \emph{exactly} (by construction) the multi-symplectic form conservation law \eqref{eq:cons} on the discrete level. The particular application to linear wave equations gave interesting results in \cite{Frank2006}. Furthermore, a numerical comparison of symplectic and multi-symplectic schemes applied to the KdV equation and its solitonic gas \cite{Dutykh2014d, Carbone2016} was undertaken in \cite{Ascher2004, Ascher2005, Dutykh2013a}. A comparative study in similar terms for the MS \textsc{Boussinesq} systems will be performed in a second, forthcoming paper.


\bigskip\bigskip
\subsection*{Acknowledgments}
\addcontentsline{toc}{subsection}{Acknowledgments}

A. D. was supported by Junta de \textsc{Castilla y Leon} and Fondos \textsc{FEDER} under the Grant VA041P17. D.M. work was supported by the \textsc{Marsden} Fund administered by the Royal Society of \textsc{New Zealand} with contract number VUW1418. A.D. would like to acknowledge the hospitality of LAMA UMR \#5127 (University \textsc{Savoie Mont Blanc}) during his visit in March 2018. We thank the reviewers for very useful comments and suggestions.


\bigskip\bigskip
\addcontentsline{toc}{section}{References}
\bibliographystyle{abbrv}
\bibliography{biblio}

\begin{thebibliography}{10}

\bibitem{Ascher2004}
U.~M. Ascher and R.~I. McLachlan.
\newblock {Multisymplectic box schemes and the Korteweg-de Vries equation}.
\newblock {\em Applied Numerical Mathematics}, 48(3-4):255--269, 2004.

\bibitem{Ascher2005}
U.~M. Ascher and R.~I. McLachlan.
\newblock {On Symplectic and Multisymplectic Schemes for the KdV Equation}.
\newblock {\em J. Sci. Comput.}, 25(1):83--104, 2005.

\bibitem{Basdevant2007}
J.-L. Basdevant.
\newblock {\em {Variational Principles in Physics}}.
\newblock Springer-Verlag, New York, 2007.

\bibitem{BBB}
T.~B. Benjamin, J.~L. Bona, and D.~K. Bose.
\newblock {Solitary-Wave Solutions of Nonlinear Problems}.
\newblock {\em Phil. Trans. R. Soc. A}, 331(1617):195--244, jun 1990.

\bibitem{BonaChK}
J.~L. Bona, H.~Chen, and O.~Karakashian.
\newblock {Stability of Solitary-Wave Solutions of Systems of Dispersive
  Equations}.
\newblock {\em Appl. Math. Optim.}, 75(1):27--53, feb 2017.

\bibitem{BCS}
J.~L. Bona, M.~Chen, and J.-C. Saut.
\newblock {Boussinesq equations and other systems for small-amplitude long
  waves in nonlinear dispersive media. I: Derivation and linear theory}.
\newblock {\em J. Nonlinear Sci.}, 12:283--318, 2002.

\bibitem{Bona2004}
J.~L. Bona, M.~Chen, and J.-C. Saut.
\newblock {Boussinesq equations and other systems for small-amplitude long
  waves in nonlinear dispersive media: II. The nonlinear theory}.
\newblock {\em Nonlinearity}, 17:925--952, 2004.

\bibitem{BCL}
J.~L. Bona, T.~Colin, and D.~Lannes.
\newblock {Long wave approximations for water waves}.
\newblock {\em Arch. Rational Mech. Anal.}, 178:373--410, 2005.

\bibitem{Bona2007}
J.~L. Bona, V.~A. Dougalis, and D.~E. Mitsotakis.
\newblock {Numerical solution of KdV-KdV systems of Boussinesq equations: I.
  The numerical scheme and generalized solitary waves}.
\newblock {\em Mat. Comp. Simul.}, 74:214--228, 2007.

\bibitem{BLS2008}
J.~L. Bona, D.~Lannes, and J.-C. Saut.
\newblock {Asymptotic models for internal waves}.
\newblock {\em J. Math. Pures Appl.}, 89(6):538--566, jun 2008.

\bibitem{Bona1975a}
J.~L. Bona and R.~Smith.
\newblock {The Initial-Value Problem for the Korteweg-De Vries Equation}.
\newblock {\em Phil. Trans. R. Soc. A}, 278(1287):555--601, jul 1975.

\bibitem{Boussinesq1877}
J.~V. Boussinesq.
\newblock {Essai sur la th{\'{e}}orie des eaux courantes}.
\newblock {\em M{\'{e}}moires pr{\'{e}}sent{\'{e}}s par divers savants {\`{a}}
  l'Acad. des Sci. Inst. Nat. France}, XXIII:1--680, 1877.

\bibitem{Bridges1997}
T.~J. Bridges.
\newblock {Multi-symplectic structures and wave propagation}.
\newblock {\em Math. Proc. Camb. Phil. Soc.}, 121(1):147--190, jan 1997.

\bibitem{Bridges1999}
T.~J. Bridges and G.~Derks.
\newblock {Unstable eigenvalues and the linearization about solitary waves and
  fronts with symmetry}.
\newblock {\em Proc. R. Soc. A}, 455(1987):2427--2469, jul 1999.

\bibitem{Bridges2001}
T.~J. Bridges and S.~Reich.
\newblock {Multi-symplectic integrators: numerical schemes for Hamiltonian PDEs
  that conserve symplecticity}.
\newblock {\em Phys. Lett. A}, 284(4-5):184--193, 2001.

\bibitem{Brocchini2013}
M.~Brocchini.
\newblock {A reasoned overview on Boussinesq-type models: the interplay between
  physics, mathematics and numerics}.
\newblock {\em Proc. R. Soc. A}, 469(2160):20130496, oct 2013.

\bibitem{Carbone2016}
F.~Carbone, D.~Dutykh, and G.~A. El.
\newblock {Macroscopic dynamics of incoherent soliton ensembles: Soliton gas
  kinetics and direct numerical modelling}.
\newblock {\em EPL}, 113(3), 2016.

\bibitem{Champneys}
A.~R. Champneys.
\newblock {Homoclinic orbits in reversible systems and their applications in
  mechanics, fluids and optics}.
\newblock {\em Phys. D}, 112(1-2):158--186, jan 1998.

\bibitem{BonaCh}
H.~Chen and J.~Bona.
\newblock {Solitary waves in nonlinear dispersive systems}.
\newblock {\em Discrete Contin. Dynam. Syst. Ser. B}, 2(3):313--378, may 2002.

\bibitem{Chen2002}
J.-B. Chen, M.-Z. Qin, and Y.-F. Tang.
\newblock {Symplectic and multi-symplectic methods for the nonlinear
  Schr{\"{o}}dinger equation}.
\newblock {\em Computers {\&} Mathematics with Applications},
  43(8-9):1095--1106, apr 2002.

\bibitem{Chen2011}
Y.~Chen, S.~Song, and H.~Zhu.
\newblock {The multi-symplectic Fourier pseudospectral method for solving
  two-dimensional Hamiltonian PDEs}.
\newblock {\em J. Comp. Appl. Math.}, 236(6):1354--1369, oct 2011.

\bibitem{Chhay2016}
M.~Chhay, D.~Dutykh, and D.~Clamond.
\newblock {On the multi-symplectic structure of the Serre-Green-Naghdi
  equations}.
\newblock {\em J. Phys. A: Math. Gen}, 49(3):03LT01, jan 2016.

\bibitem{Christov2001}
C.~I. Christov.
\newblock {An energy-consistent dispersive shallow-water model}.
\newblock {\em Wave Motion}, 34:161--174, 2001.

\bibitem{Clamond2012b}
D.~Clamond and D.~Dutykh.
\newblock {Fast accurate computation of the fully nonlinear solitary surface
  gravity waves}.
\newblock {\em Comput. {\&} Fluids}, 84:35--38, jun 2013.

\bibitem{Clamond2016a}
D.~Clamond and D.~Dutykh.
\newblock {Multi-symplectic structure of fully nonlinear weakly dispersive
  internal gravity waves}.
\newblock {\em J. Phys. A: Math. Gen.}, 49(31):31LT01, aug 2016.

\bibitem{Daripa2006}
P.~Daripa.
\newblock {Higher-order Boussinesq equations for two-way propagation of shallow
  water waves}.
\newblock {\em Eur. J. Mech. B/Fluids}, 25(6):1008--1021, nov 2006.

\bibitem{DeDonder1930}
T.~de~Donder.
\newblock {\em {Th{\'{e}}orie invariantive du calcul des variations}}.
\newblock Gauthier-Villars, Paris, 1930.

\bibitem{DMII}
V.~A. Dougalis and D.~E. Mitsotakis.
\newblock {Theory and numerical analysis of Boussinesq systems: A review}.
\newblock In N.~A. Kampanis, V.~A. Dougalis, and J.~A. Ekaterinaris, editors,
  {\em Effective Computational Methods in Wave Propagation}, pages 63--110. CRC
  Press, 2008.

\bibitem{Dutykh2013a}
D.~Dutykh, M.~Chhay, and F.~Fedele.
\newblock {Geometric numerical schemes for the KdV equation}.
\newblock {\em Comp. Math. Math. Phys.}, 53(2):221--236, 2013.

\bibitem{Dutykh2013b}
D.~Dutykh and D.~Clamond.
\newblock {Efficient computation of steady solitary gravity waves}.
\newblock {\em Wave Motion}, 51(1):86--99, jan 2014.

\bibitem{Dutykh2010e}
D.~Dutykh, T.~Katsaounis, and D.~Mitsotakis.
\newblock {Finite volume methods for unidirectional dispersive wave models}.
\newblock {\em Int. J. Num. Meth. Fluids}, 71:717--736, 2013.

\bibitem{Dutykh2014d}
D.~Dutykh and E.~Pelinovsky.
\newblock {Numerical simulation of a solitonic gas in KdV and KdV-BBM
  equations}.
\newblock {\em Phys. Lett. A}, 378(42):3102--3110, aug 2014.

\bibitem{Frank2006}
J.~Frank, B.~E. Moore, and S.~Reich.
\newblock {Linear PDEs and Numerical Methods That Preserve a Multisymplectic
  Conservation Law}.
\newblock {\em SIAM J. Sci. Comput.}, 28(1):260--277, jan 2006.

\bibitem{Friedrichs1971}
K.~O. Friedrichs and P.~D. Lax.
\newblock {Systems of Conservation Equations with a Convex Extension}.
\newblock {\em PNAS}, 68(8):1686--1688, 1971.

\bibitem{GearG}
J.~A. Gear and R.~Grimshaw.
\newblock {Weak and Strong Interactions between Internal Solitary Waves}.
\newblock {\em Stud. Appl. Math.}, 70(3):235--258, jun 1984.

\bibitem{Goldschmidt1973}
H.~Goldschmidt and S.~Sternberg.
\newblock {The Hamilton-Cartan formalism in the calculus of variations}.
\newblock {\em Ann. Inst. Fourier}, 23(1):203--267, 1973.

\bibitem{Grillakis1987}
M.~Grillakis, J.~Shatah, and W.~Strauss.
\newblock {Stability theory of solitary waves in the presence of symmetry, I}.
\newblock {\em Journal of Functional Analysis}, 74(1):160--197, sep 1987.

\bibitem{Grillakis1990}
M.~Grillakis, J.~Shatah, and W.~Strauss.
\newblock {Stability theory of solitary waves in the presence of symmetry, II}.
\newblock {\em Journal of Functional Analysis}, 94(2):308--348, dec 1990.

\bibitem{Hakkaev}
S.~Hakkaev.
\newblock {Stability and instability of solitary wave solutions of a nonlinear
  dispersive system of Benjamin-Bona-Mahony type}.
\newblock {\em Serdica Math. J.}, 29:337--354, 2003.

\bibitem{Huang2003}
L.-Y. Huang, W.-P. Zeng, and M.-Z. Qin.
\newblock {A new multi-symplectic scheme for nonlinear "good" Boussinesq
  equation}.
\newblock {\em J. Comp. Math.}, 21(6):703--714, 2003.

\bibitem{IoosA}
G.~Iooss and M.~Adelmeyer.
\newblock {\em {Topics in Bifurcation Theory and Applications}}.
\newblock World Scientific, Singapore, 2 edition, 1999.

\bibitem{IoosK}
G.~Iooss and K.~Kirchg{\"{a}}ssner.
\newblock {Water waves for small surface tension: an approach via normal form}.
\newblock {\em Proc. R. Soc. Edinburgh Sect. A}, 122(3-4):267--299, nov 1992.

\bibitem{Islas2003}
A.~L. Islas and C.~M. Schober.
\newblock {Multi-symplectic Spectral Methods for the Sine-Gordon Equation}.
\newblock In P.~M.~A. Sloot, D.~Abramson, A.~V. Bogdanov, Y.~E. Gorbachev,
  J.~J. Dongarra, and A.~Y. Zomaya, editors, {\em Computational Science - ICCS
  2003}, pages 101--110. Springer, Berlin, Heidelberg, 2003.

\bibitem{Ke2015}
Q.~Ke, E.~Ferrara, F.~Radicchi, and A.~Flammini.
\newblock {Defining and identifying Sleeping Beauties in science}.
\newblock {\em Proceedings of the National Academy of Sciences},
  112(24):7426--7431, jun 2015.

\bibitem{Kijowki1974}
J.~Kijowki.
\newblock {Multiphase spaces and gauge in calculus of variations}.
\newblock {\em Bull. Acad. Polon. des Sci., S{\'{e}}rie Sci. Math., Astr. et
  Phys.}, XXII:1219--1225, 1974.

\bibitem{Krupka1975}
D.~Krupka.
\newblock {A geometric theory of ordinary first order variational problems in
  fibered manifolds. I. Critical sections}.
\newblock {\em Journal of Mathematical Analysis and Applications},
  49(1):180--206, jan 1975.

\bibitem{Krupka1975a}
D.~Krupka.
\newblock {A geometric theory of ordinary first order variational problems in
  fibered manifolds. II. Invariance}.
\newblock {\em Journal of Mathematical Analysis and Applications},
  49(2):469--476, feb 1975.

\bibitem{Lagrange1853}
J.-L. Lagrange.
\newblock {\em {M{\'{e}}canique analytique}}.
\newblock Hallet-Bachelier, Paris, 3 edition, 1853.

\bibitem{Leimkuhler2004}
B.~Leimkuhler and S.~Reich.
\newblock {\em {Simulating Hamiltonian Dynamics}}, volume~14 of {\em Cambridge
  Monographs on Applied and Computational Mathematics}.
\newblock Cambridge University Press, Cambridge, 2005.

\bibitem{Lepage1936}
T.~Lepage.
\newblock {Sur les champs g{\'{e}}od{\'{e}}siques du calcul des variations}.
\newblock {\em Bull. Acad. Roy. Belg., Cl. Sci}, 27:716--729, 1036--1046, 1936.

\bibitem{Li2013}
H.~Li and J.~Sun.
\newblock {A new multi-symplectic Euler box scheme for the BBM equation}.
\newblock {\em Math. Comp. Model.}, 58(7-8):1489--1501, oct 2013.

\bibitem{Lions1984}
P.-L. Lions.
\newblock {The concentration-compactness principle in the calculus of
  variations. The locally compact case, part 1}.
\newblock {\em Annales de l'I.H.P. Analyse non lin{\'{e}}aire}, 1(2):109--145,
  1984.

\bibitem{Lions1984a}
P.-L. Lions.
\newblock {The concentration-compactness principle in the calculus of
  variations. The locally compact case, part 2}.
\newblock {\em Annales de l'I.H.P. Analyse non lin{\'{e}}aire}, 1(4):223--283,
  1984.

\bibitem{Lombardi2000}
E.~Lombardi.
\newblock {\em {Oscillatory integrals and phenomena beyond all algebraic orders
  with applications to homoclinic orbits in reversible systems}}.
\newblock Springer, 2000.

\bibitem{Madsen1999}
P.~A. Madsen and H.~A. Schaffer.
\newblock {A review of Boussinesq-type equations for surface gravity waves}.
\newblock {\em Adv. Coastal Ocean Engng}, 5:1--94, 1999.

\bibitem{Marsden1998}
J.~E. Marsden, G.~W. Patrick, and S.~Shkoller.
\newblock {Multisymplectic geometry, variational integrators, and nonlinear
  PDEs}.
\newblock {\em Comm. Math. Phys.}, 199(2):351--395, 1998.

\bibitem{Moore2003a}
B.~Moore and S.~Reich.
\newblock {Multi-symplectic integration methods for Hamiltonian PDEs}.
\newblock {\em Future Generation Computer Systems}, 19(3):395--402, 2003.

\bibitem{Peregrine1967}
D.~H. Peregrine.
\newblock {Long waves on a beach}.
\newblock {\em J. Fluid Mech.}, 27:815--827, 1967.

\bibitem{Souriau1997}
J.-M. Souriau.
\newblock {\em {Structure of Dynamical Systems: a Symplectic View of Physics}}.
\newblock Birkh{\"{a}}user, Boston, MA, 1997.

\bibitem{Spivak1971}
M.~Spivak.
\newblock {\em {Calculus on Manifolds: A Modern Approach to Classical Theorems
  of Advanced Calculus}}.
\newblock Westview Press, Princeton, 1971.

\bibitem{Sun2004}
Y.-J. Sun and M.-Z. Qin.
\newblock {A multi-symplectic scheme for RLW equation}.
\newblock {\em J. Comp. Math.}, 22(4):611--621, 2004.

\bibitem{Toland}
J.~F. Toland.
\newblock {Existence of symmetric homoclinic orbits for systems of
  Euler-Lagrange equations}.
\newblock {\em AMS Proc. Symposia in Pure Math.}, 45(2):447--459, 1986.

\bibitem{Volterra1890a}
V.~Volterra.
\newblock {Sopra una estensione della teoria Jacobi-Hamilton del calcolo delle
  variazioni}.
\newblock {\em Rend. Cont. Acad. Lincei, ser. IV}, VI:127--138, 1890.

\bibitem{Volterra1890}
V.~Volterra.
\newblock {Sulle equazioni differenziali che provengono da questiono di calcolo
  delle variazioni}.
\newblock {\em Rend. Cont. Acad. Lincei, ser. IV}, VI:42--54, 1890.

\bibitem{Weinstein1986}
M.~I. Weinstein.
\newblock {Lyapunov stability of ground states of nonlinear dispersive
  evolution equations}.
\newblock {\em Comm. Pure Appl. Math.}, 39(1):51--67, jan 1986.

\bibitem{Weyl1935}
H.~Weyl.
\newblock {Geodesic Fields in the Calculus of Variation for Multiple
  Integrals}.
\newblock {\em Annals of Mathematics}, 36(3):607--629, jul 1935.

\bibitem{Zhao2000}
P.~F. Zhao and M.~Z. Qin.
\newblock {Multisymplectic geometry and multisymplectic Preissmann scheme for
  the KdV equation}.
\newblock {\em J. Phys. A: Math. Gen.}, 33(18):3613--3626, may 2000.

\end{thebibliography}
\bigskip\bigskip

\end{document}